%% file: praffai_etal_2013.tex
\documentclass[12pt]{iopart}
\usepackage{iopams}
\usepackage{graphicx}
\usepackage{amssymb}
\usepackage{multirow}
\usepackage{subfig}
\usepackage{color}

\begin{document}

\title{Optimal Networks of Future Gravitational-Wave Telescopes}

\author{P\'eter Raffai$^{1,2,3}$, L\'aszl\'o Gond\'an$^{2,3}$, Ik Siong Heng$^4$, N\'andor Kelecs\'enyi$^{2,3}$, Josh Logue$^4$, Zsuzsa M\'arka$^1$, Szabolcs M\'arka$^1$}
\address{$^1$ Columbia University, Department of Physics, New York, NY 10027, USA\\}
\address{$^2$ E\"otv\"os University, Institute of Physics, 1117 Budapest, Hungary\\}
\address{$^3$ MTA-ELTE EIRSA "Lend\"ulet" Astrophysics Research Group, 1117 Budapest, Hungary\\}
\address{$^4$ SUPA, University of Glasgow, Glasgow, G12 8QQ, United Kingdom\\}
\ead{praffai@bolyai.elte.hu}

\begin{abstract}
We aim to find the optimal site locations for a hypothetical network of $1-3$ triangular gravitational-wave telescopes. We define the following $N$-telescope figures of merit (FoMs) and construct three corresponding metrics: (a) capability of reconstructing the signal polarization; (b) accuracy in source localization; and (c) accuracy in reconstructing the parameters of a standard binary source. We also define a combined metric that takes into account the three FoMs with practically equal weight. After constructing a geomap of possible telescope sites, we give the optimal $2$-telescope networks for the four FoMs separately in example cases where the location of the first telescope has been predetermined. We found that based on the combined metric, placing the first telescope to Australia provides the most options for optimal site selection when extending the network with a second instrument. We suggest geographical regions where a potential second and third telescope could be placed to get optimal network performance in terms of our FoMs. Additionally, we use a similar approach to find the optimal location and orientation for the proposed LIGO-India detector within a five-detector network with Advanced LIGO (Hanford), Advanced LIGO (Livingston), Advanced Virgo, and KAGRA. We found that the FoMs do not change greatly in sites within India, though the network can suffer a significant loss in reconstructing signal polarizations if the orientation angle of an L-shaped LIGO-India is not set to the optimal value of $\sim 58.2^\circ(+k\times 90^\circ)$ (measured counterclockwise from East to the bisector of the arms).
\end{abstract}

\submitto{\CQG}

\maketitle

\section{Introduction}
\label{section_introduction}
\input{Introduction}

\section{Selection of possible telescope sites}
\label{section_map}
\input{map}

\section{$N$-telescope FoMs and metrics}
\label{section_maths}
\input{math_background}

\section{Optimal networks of $\Delta$-telescopes}
\label{section_positions}
\input{opt_position}

\section{The LIGO-India case in the second generation GW detection network}
\label{section_India}
\input{opt_India}

\section{Discussion}
\label{section_discussion}
\input{Discussion}

\ack
\label{section_acknowledgments}
\input{Acknowledgments}

\section*{References}
\bibliographystyle{unsrt}
\bibliography{References_FutureNetPaper}

\end{document}

%% file: Introduction.tex
Modern gravitational-wave (GW) detectors are state-of-the-art interferometers with arm lengths typically on the kilometer scale. The most advanced network of such instruments so far consists of the two detectors of the Laser Interferometer Gravitational-wave Observatory (LIGO) \cite{BPAetal09} at Hanford (Washington, USA) and at Livingston (Louisiana, USA), both having $4$ km long arms, and the Virgo detector at Cascina (Italy) \cite{TAetal12}, having a $3$ km arm length. All three detectors are currently undergoing major upgrades until they will come back online around 2014-2017 as a network of second generation GW detectors, named as Advanced LIGO \cite{AdvLIGO} and Advanced Virgo \cite{AdvVIRGO}, respectively.

Two additional L-shaped detectors are proposed to become operational shortly after 2018, and will join the network of second generation detectors: the LIGO-India in India \cite{LIGOIndia}, and the KAGRA detector in Japan \cite{KAGRA}, with $4$ and $3$ km long arms, respectively. The KAGRA detector is to be constructed in the Kamioka mine, which already sets the location and orientation of this future instrument. However, the site location and the orientation of the arms of LIGO-India have not yet been finalized. This provides an opportunity to suggest both based on purely scientific figures of merit (FoMs).

According to source models, the detection of GWs is expected shortly after the second generation GW detectors start taking data with their proposed nominal sensitivity \cite{JA10}. This opens up the era of {\it GW astronomy}, where the expected common detection of GWs provides an opportunity for observing distant sources and yet unexplored physical processes, making GW detectors powerful tools of astronomical research, similar to electromagnetic telescopes. To emphasize this major shift in the functionality of these kilometer-scale interferometers, we often use the term {\it GW telescopes} to refer to future GW detectors during and beyond the second generation era. Since the results in this paper correspond to such future instruments, we are going to use the term {\it GW telescopes} to refer to them, and keep the term {\it GW detectors} for instruments that primarily aim to make the first detection of GWs. In this terminology, advanced (second generation) interferometers should only be categorized as GW telescopes after they have undergone the expected shift from the first detection to the routine detection era.

Precision GW astronomy requires GW telescopes even beyond the expected sensitivities of second generation interferometers, extending also at low frequencies. The Einstein Telescope (ET) project \cite{ET} is a good example that aims to pioneer novel design configurations, construction methods, and operational techniques, to overcome the limitations of advanced interferometers in GW astronomy. The final goal of the ET project is to provide a baseline for the future construction of a {\it third generation} GW telescope. This, according to the conceptual document \cite{MA11}, will consist of three interferometers (single-interferometer design) or interferometer {\it pairs} (xylophone design) with $10$ km long arms, built underground in a way to form an equilateral triangle shape. The ET is aspiring to have a noise level about one order of magnitude below the noise floor of Advanced LIGO in its most sensitive range, and extend sensitivity at lower frequencies \cite{MP10}. This would make the ET capable of observing GW sources at distances of even up to several gigaparsecs, depending on the source type \cite{MA11}.

GW telescopes are not equally sensitive to GWs coming from different sky directions. Beyond the scientific advantages of improved sensitivity and of having multiple interferometers at one telescope site, the equilateral triangle geometry makes the directional sensitivity of such a GW telescope cylindrically symmetric around the symmetry axis of the triangle for both GW polarizations \cite{AF09}. In practice, this means that all the scientific FoMs we can define in terms of GW detection will be independent from how the triangle is oriented at a given geographical position. For a network of such GW telescopes, the corresponding FoMs will only depend on the geographical positions of the individual GW telescopes relative to each other.

\begin{figure}
    \centering
    \includegraphics[width=100mm]{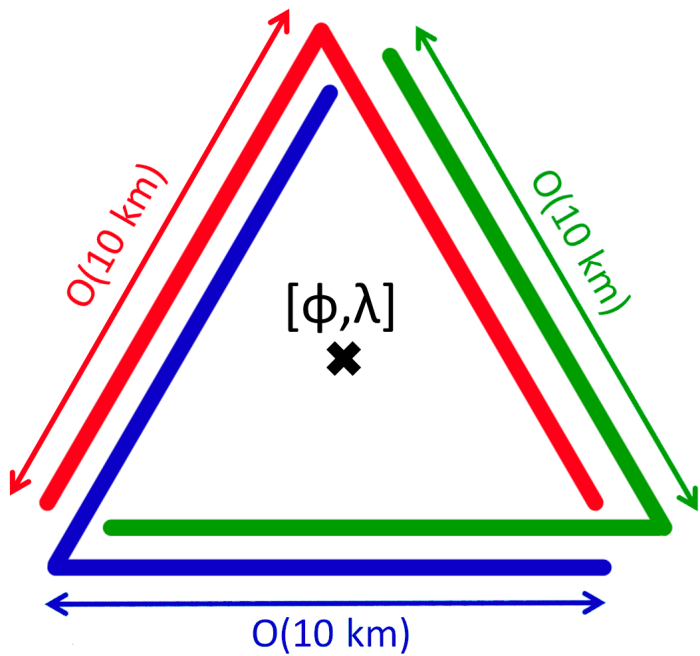}
\caption{The sketch of a hypothetical {\it Delta telescope} ($\Delta$-telescope). The instrument consists of three identical interferometers (single-interferometer design) or interferometer pairs (xylophone design) with $\cal{O}$$(10\ \mathrm{km})$ long arms. The opening angle between the arms of each interferometer is $60^{\circ}$, and the interferometers are oriented such that they form an equilateral triangle shape. The center of the triangle (marked with $\times$) is placed to a geographical position represented by $[\Phi,\lambda]$ latitude and longitude coordinates, respectively. Note, that this is the baseline design geometry proposed by the ET design document \cite{MA11}.}
\label{fig:Delta_telescope}
\end{figure}

The scientific advantages of the geometrical design proposed by the ET project sets a long term model for future GW telescope development in the third-generation era and possibly even beyond. Therefore to preserve the generality of our discussions, from now on we will use the term {\it Delta telescope} ($\Delta$-telescope) to refer to a future or hypothetical GW telescope that consists of three $\cal{O}$$(10\ \mathrm{km})$ scale interferometers sharing the same site and forming an equilateral triangle shape, similarly to the proposed design for the ET. The sketch of a $\Delta$-telescope is shown in figure \ref{fig:Delta_telescope}.

Although there are some promising site candidates based on local noise background measurements \cite{MA11}, the future geographical position of the first $\Delta$-telescope is still unknown. Beyond the expected local noise background, one can also consider FoMs of a {\it network} of GW telescopes when selecting the best site candidates for any number of future telescopes. In doing so, we should first consider what type of GW telescopes we should take into account in a future network that will include at least one $\Delta$-telescope.

We consider a first $\Delta$-telescope that provides an improvement over the second-generation GW telescopes in the limiting sensitivity by more than a factor of $10$ over their entire frequency band \cite{MP10}. This yields a far better detector performance, and different physics to be explored \cite{MP10,ECM11}. As such an instrument becomes operational (after $2020$ at earliest \cite{MA11}), the second-generation GW telescopes with less than $10\%$ sensitivity and $0.1\%$ explorable volume will become irrelevant as coherent detection partners. Therefore we only consider $N$-telescope networks here where {\it all} the nodes are $\Delta$-telescopes, when exploring the best site locations based on $N$-telescope FoMs.

In case of a single $\Delta$-telescope, the first FoM that we need to take into account when choosing the site location from an allowed set of locations is the expected noise background caused by local environmental effects. However, when building a network of more than one $\Delta$-telescope, the selection of the first site already severely restricts the optimal location for the second (and third, fourth, etc.) telescope, based on the $N$-telescope FoMs. By not taking this into account when selecting the site for the first telescope, one might get into a situation where the set of site candidates for the additional detectors has already narrowed down to options that are sub-optimal in terms of the $N$-telescope FoMs. Therefore even though there has not been any published plans yet to construct more than one $\Delta$-telescopes, it is critical to keep the door open for a future extension to $N>1$ $\Delta$-telescopes and explore the best site candidates for a network of multiple $\Delta$-telescopes. Thus, we emphasize that the results based on $N$-telescope FoMs should be taken into account in the site selection of the first $\Delta$-telescope, which we treat as the {\it primary} element constructed in a possible future network of GW telescopes.

In this paper, we suggest optimal locations for $1-3$ $\Delta$-telescopes from a set of possible sites chosen with respect to basic geographic limitations (e.g. availability of solid ground; elevation of the site; etc.) and expected noise levels from local environmental effects (such as near-coastal microseism and human activity). In the analysis, we intentionally disregard geopolitical considerations to allow for best science type optimization. The optimization is carried out in terms of three different $N$-telescope FoMs, and in terms of a combined example metric that takes into account the three FoMs with practically equal weight. In the optimization, for simplicity, we assume that all $\Delta$-telescopes in the network are identical and have the same sensitivity curve.

Additionally, we use the same approach to find the optimal location and orientation for the proposed LIGO-India detector within a five-detector network with Advanced LIGO (Hanford), Advanced LIGO (Livingston), Advanced Virgo, and KAGRA. In this case, again for simplicity, we assumed that these L-shaped detectors have practically the same noise levels. In the frequency band around $100$ Hz where these detectors will have their lowest level of noise, this approximation is reasonably accurate (e.g. see figure 5. in \cite{MA11}).

The paper is organized as follows. In section \ref{section_map} we discuss how the set of acceptable site locations were chosen from a geographical map of the world. In section \ref{section_maths} we introduce three $N$-telescope FoMs together with the corresponding metrics, and a combined metric defined in the three-dimensional metric space. In section \ref{section_positions}, we give the optimal $2$-telescope networks for the four FoMs separately in example cases where the location of the first telescope has been predetermined. We also suggest geographical regions in section \ref{section_positions} where the first three telescope in a network should be placed to get optimal network performance in terms of our FoMs, and to provide the most options for optimal site selection when extending the network with an additional instrument. Section \ref{section_India} discusses the results of the site location and orientation optimization for the future LIGO-India detector. Finally, we summarize and discuss our results in section \ref{section_discussion}.

%% file: map.tex
In the process of constructing a set of possible site locations for a future GW telescope, we filtered out geographical positions where either the construction of kilometer-scale interferometers would be unfeasible, or where we expect to have permanent local noise sources that would strongly limit the scientific performance of the telescope during observational runs.

We started our selection process with a cylindrically projected map of the world, where the oceanic surfaces have already been filtered out. We removed the polar regions from the map due to the harsh weather conditions that would make the construction and operation of a GW telescope unfeasible. We also subtracted all the continental lakes from our map. Finally, we applied a cut at an elevation limit of $2000$ meters above sea level, due to the limited accessibility of such places.

We applied two additional filters that were based on considerations of local noise sources. Due to the high level of excess near field seismic noise caused by oceanic waves clashing to the continental shelf, we removed a $\sim 100$ km wide zone along the coastlines of continents and islands. To exclude areas that are permanently affected by intense human activity that leads to an increased level of gravity gradient background, we used a high-resolution image of Earth's city lights from the Visible Earth Catalog of the National Aeronautics and Space Administration (NASA) \cite{NASAVE}. By removing the high-intensity pixels of the city lights map from our map of site candidates, we excluded both the urban areas and the most heavily used traffic roads.

\begin{figure}
    \centering
    \includegraphics[width=150mm]{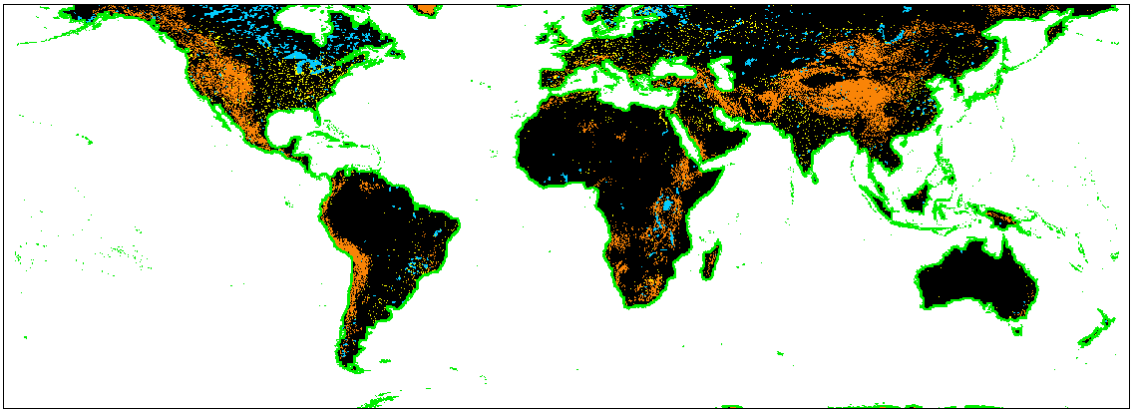}
\caption{The map of acceptable regions for the construction of a future $\Delta$-telescope (black), together with the excluded geographical areas: oceanic surfaces (white), continental lakes (blue), locations at elevations higher than $2000\ \mathrm{m}$ above sea level (brown), near-coastal regions within $\sim 100$ kilometers (green), and urban areas (yellow). The remaining locations (black) were discretely sampled and were used as a set of acceptable telescope sites for our numerical analysis.}
\label{fig:map}
\end{figure}

The resulting set of acceptable sites, together with the excluded geographical areas are visualized in figure \ref{fig:map}. The remnant geographical areas were discretely sampled to get a sub-sample of locations where the geographical positions are $\sim 300$ km away from each other. The resulting $1525$ locations were treated as the set of acceptable sites for $\Delta$-telescopes. Note, that future analyses with additional constraints and refinements might nuance the local selection of acceptable sites, but will not affect the global picture we are about to present here.

%% file: math_background.tex
In this section we introduce the FoMs that we used to evaluate the scientific potential of the different configurations of a $\Delta$-telescope network. Our framework in many ways is similar to the ones introduced in \cite{BFS11} and in \cite{RW10}. A similar but alternative set of FoMs was also suggested in \cite{ACS06}, where extending the network of initial GW-detectors was considered.

As discussed in section \ref{section_introduction}, the FoMs for a network of $\Delta$-telescopes are independent from the orientation of the individual telescopes. We also pointed out that the horizon distances of $\Delta$-telescopes for the targeted GW source types are in the order of many gigaparsecs, in which the distribution of matter can be approximated as homogeneous and isotropic. Due to the rotation of Earth and its orbital motion around the Sun, the directional sensitivity of a ground-based GW telescope changes with time towards the different directions in the galactic frame. By taking into account all these effects, as a first approximation, we only need to consider the positions of the GW telescopes in the network {\it relative} to each other, and use all-sky averaged FoMs to characterize the network.

As an alternative, one can configure a network such as to maximize the observational time with the highest directional sensitivity towards a certain sky direction. This sky direction could be chosen based on known inhomogeneities or anisotropies in the mass distribution of the local Universe. However, in this work we optimized the network configurations based on all-sky averaged FoMs, leaving the local mass distribution as the scope of a follow-up paper.

In the following subsections, we are going to introduce three all-sky averaged FoMs together with the corresponding metrics to characterize the different configurations of $\Delta$-telescope networks. Additionally, we are going to discuss why it is unnecessary to consider the average directional sensitivity (or {\it reach} \cite{BFS11}) of the network as a fourth FoM. We summarize our FoMs and metrics at the end of this section, in table~\ref{Table:metrics}.

\subsection{Capability of reconstructing the signal polarization}
\label{subsect:I}

Interferometric GW telescopes sense the change in the distance between their test masses along the lines of the interferometer arms. Therefore in order to model the strain output of a telescope for a GW signal arriving from a given sky direction, one must project the GW strain to the reference frame set by the interferometer arms. The projection can be given as a linear combination of the $+$ and $\times$ polarization components of the incoming GW signal, where the linear factors, $F_+$ and $F_{\times}$ (ranging from 0 to 1, and called ''antenna factors'') depend on the sky direction relative to the interferometer arms, and on the angle that sets the orientation of the frame in which the two polarizations are defined. This frame is called the ''polarization frame'', while the corresponding angle is called the ''polarization angle'', and conventionally denoted by $\Psi$.

By choosing a $\Psi$ polarization angle, one can plot the antenna factors for any ground-based interferometers as functions of the different sky directions given by $\Phi$ latitude and $\lambda$  longitude coordinates in the Earth-centered coordinate system. These two-dimensional plots are called ''antenna patterns'', and they can be constructed for a network of $N$ $\Delta$-telescopes by using the same polarization angle for all interferometers, and calculating the root-mean-square average of the $+$ and $\times$ antenna factors for the $3N$ individual interferometers:

\begin{equation}\label{eq:Fnetwork}
F^\mathrm{network} = \sqrt{\frac{F_1^2+F_2^2+F_3^2+...+F_{3N}^2}{3N}}.
\end{equation}

Here the $F$ antenna factors can either all correspond to the $+$ or to the $\times$ signal polarization. Also, for any number of interferometers and for any sky directions one can always choose the polarization angle such that $F_+$ is maximized, and $F_{\times}$ is minimized, and therefore $F_+\geq F_{\times}$. This special polarization frame is called the Dominant Polarization Frame (DPF) (see \cite{SK11} and \cite{Dragothesis} for more details). Normally, when constructing the DPF for multiple interferometers, one needs to take into account their noise levels relative to each other in the frequency band of interest \cite{Dragothesis}. However when all interferometers are identical (as in the case of $N$ $\Delta$-telescopes), this consideration is not necessary.

The sensitivity of a network of telescopes towards a chosen direction is usually characterized by the root-mean-square average of $F^\mathrm{network}_+$ and $F^\mathrm{network}_{\times}$. However, depending on the network configuration, the ratio of $F^\mathrm{network}_{\times}$ and $F^\mathrm{network}_+$ within this average can have any value from zero to one (note that $F^\mathrm{network}_+\geq F^\mathrm{network}_{\times}$ is always true in the DPF). This ratio towards a given sky direction characterizes the relative network sensitivity to the two GW polarizations, and is conventionally called the {\it network alignment factor} \cite{SK11}. The value of this factor determines the ratio of the signal-to-noise ratios from the two GW polarization components, assuming that in average their sum-square energies are the same.

Having a $F^\mathrm{network}_{\times}/F^\mathrm{network}_+$ ratio equal to $1$ in the DPF means that the network is equally sensitive to both polarizations of a GW signal incoming from the given direction. On the other hand, if the $F^\mathrm{network}_{\times}/F^\mathrm{network}_+$ ratio is significantly lower than $1$ in the DPF, that means that the network will be relatively insensitive to the $\times$ polarization component of the signal ($\times$ being defined in the DPF), and thus the network will have a limited capability of detecting the $\times$ component of a GW signal incoming from the given direction. Since in order to reconstruct the polarization of a GW signal, we need to be able to detect and reconstruct the energy content of the signal for both polarizations, $F^\mathrm{network}_{\times}/F^\mathrm{network}_+\approx 1$ can be interpreted as optimal in reconstructing the signal polarization for the given sky direction, provided that the overall signal strength is sufficient. Therefore to characterize a telescope network configuration in terms of its capability of reconstructing the signal polarization, we can use the  $|F^\mathrm{network}_+-F^\mathrm{network}_{\times}|^2$ quantity averaged over all sky directions:

\begin{equation}\label{eq:I}
I=\left (\frac{1}{4\pi}\oint \left | F^\mathrm{network}_+\left ( \Phi ,\lambda  \right ) - F^\mathrm{network}_{\times}\left ( \Phi ,\lambda  \right ) \right |^2 d\Omega  \right )^{-\frac{1}{2}}
\end{equation}

where $\Phi$ and $\lambda$ are the latitude and longitude coordinates of the different sky directions defined in the Earth-centered coordinate system, and $d\Omega=\cos{\Phi}d\Phi d\lambda$ is the infinitesimal solid angle towards the [$\Phi$,$\lambda$] direction. If we want to construct an optimal network of telescopes by maximizing its capability of reconstructing the signal polarization, we should minimize the all-sky average of the $|F^\mathrm{network}_+-F^\mathrm{network}_{\times}|^2$ difference. As shown by Eq. \ref{eq:I}, the $I$ metric is defined in a way that it should be {\it maximized} instead.

\subsection{Accuracy in source localization}
\label{subsect:D}

Transient signals such as e.g. binary inspiral signals could stay in the sensitive frequency band of a future $\Delta$-telescope as long as $10$ days \cite{MP10}. In case of such signal durations, the Doppler modulation of the signal due to Earth's rotation and revolution can be used to constrain the position of the source in the sky even with a single $\Delta$-telescope \cite{MP10}. Note, however, that at a frequency of $1$ Hz the resolution of a future $\Delta$-telescope is expected to be only around $1$ str even if we consider the baseline from the Earth's motion around the Sun over $10$ days \cite{MP10}. Therefore (similarly to the case of an advanced detector network \cite{SK11}) for transient signals, an effective source localization in the sky can only be achieved by triangulation based on the times of detection of the incoming signal with multiple GW-telescopes \cite{SF11}. 

The detection times can only be measured with an accuracy that is directly proportional to the signal-to-noise ratio, and thus to one over the strain noise level of the interferometer within the bandwidth of the signal \cite{SF11}. This uncertainty in the detection time measurement can be translated into uncertainty in the source localization for the signal. However, when the goal is to compare the accuracy in source localization for different configurations of $N$ identical telescopes, the noise levels of the interferometers do not need to be taken into account when constructing a metric.

Since the triangulation is based on the {\it differences} between the detection times, the accuracy in source localization also depends on the distances between the telescope sites along the line-of-sight to the GW source. If we want to maximize the average accuracy of source localization towards all sky directions, we must place the telescopes of the network as far away from each other as possible in the three dimensional space. For a $2$-telescope network, this makes the distance between the two telescopes an obvious choice for a metric to be maximized. For a $3$-telescope network, it is the {\it area} of the triangle formed by the three telescopes, that should be maximized (e.g. see Eq.$(1)$ in \cite{LW10}). In order to have the same units for the $2$- and $3$-telescope metric, we use the {\it square} of the distance between the telescopes as the metric in the $2$-telescope case. In both cases, we denote the metric corresponding to the potential of the network in source localization as $D$.

In case of the LIGO-India tests, where the network consists of $N=5$ telescopes, we choose $D$ to be the area of the triangle with the largest area among the triangles formed by all possible combinations of three telescopes in the network. Thus, as a simplification, we characterize the $N=5$ network with the $D$ metric value of its best $N=3$ sub-network in terms of accuracy in source localization.

\subsection{Accuracy in reconstructing the parameters of a standard binary source}
\label{subsect:R}

The most promising candidates of GW emission observable by future GW telescopes are binary systems of black holes (BHs) and/or neutron stars (NSs). Based on the models of such systems and the proposed sensitivity curve for ET, future $\Delta$-telescopes should be able to detect the GW emission of these binaries up to several or several tens of gigaparsecs, depending on the total mass of the system \cite{MA11}.

The waveforms of the GW signals from these sources are known with a relatively high accuracy from post-Newtonian calculations even in cases when the spins of the binary components are not negligible \cite{CC94,CMG94}. This makes binary systems of BHs and/or NSs {\it standard sources} in the sense that they can be used to characterize the capabilities of a GW telescope or a network of GW telescopes in terms of distance reach and accuracy in source parameter reconstruction.

In case of a circular binary of spinless compact stars, the typical parameters to be reconstructed from the GW waveform are the chirp mass of the system and the distance of the source \cite{CC94}. The alignment of the orbital plane relative to the line-of-sight is reconstructed from the relative power in the $+$ and $\times$ components of the incoming signal \cite{CC94}, and thus the performance of the telescope network in terms of this is already being tested by the $I$ metric (see section \ref{subsect:I}).

If we choose a configuration of $N$ identical $\Delta$-telescopes, we can use the Fisher matrix method \cite{AA09} to calculate the chirp mass and source distance reconstruction errors for a standard GW source. The source that we chose is an inspiraling binary of spinless neutron stars, each having a mass of $1.4M_{\odot}$, placed to a distance of $1\ \mathrm{Gpc}$ from Earth, and aligned such that the orbital plane is always perpendicular to the line-of-sight. For this standard source, the sensitivity curves for the single-interferometer (ET-B in \cite{MA11}) and the xylophone designs (ET-C or ET-D in \cite{MA11}) are practically the same in the frequency band of the emitted GW signal, and thus the results are independent from the specific sensitivity curve we choose from these three options (we used ET-B in our calculations). Also, as we pointed out earlier, a future $\Delta$-telescope should have no problem detecting such a binary source at a $1\ \mathrm{Gpc}$ distance, and reconstructing its chirp mass.

Generally the orbit of an inspiraling binary is elliptical, and the principal axes of the ellipse give a preferred polarization basis \cite{CC94,JTW05}. However, when the line-of-sight is perpendicular to the orbital plane, the components of the base vectors as defined in \cite{CC94} become singular. Since the incoming GW signal is circularly polarized in our case, we can choose any two orthogonal vectors as polarization bases perpendicular to the line-of-sight, and resolve the singularity. In choosing such a polarization basis, as an approximation, we neglected the change in the direction of the line-of-sight throughout the time while the signal is within the sensitive band of the $\Delta$-telescope.

Using the waveform of the spinless binary in the Newtonian approximation \cite{CC94}, we calculated the components of the Fisher information matrix \cite{BK07} for each single interferometer, and summed it up for all the interferometers in the $\Delta$-telescope sites. We found that it is unnecessary to construct a separate metric based on chirp mass and source distance reconstruction errors, because the results are the same when we compare the different network configurations using the two metrics. Thus we chose a metric that is based on chirp mass reconstruction error only. We also found that the chirp mass reconstruction errors only slightly depend on the off-diagonal elements of the Fisher information matrix (the results with and without the off-diagonal elements are within less than $0.5\%$). Thus for simplicity, we only used the diagonal elements of the Fisher information matrix when calculating the chirp mass reconstruction errors.

The chirp mass reconstruction error, $\delta \cal{M}$, depends on the source location relative to the positions of the telescope sites ($\delta \cal{M}$$=\delta \cal{M}$$(\Phi,\lambda)$). In order to identify the best telescope network configuration, we used the all-sky average of the {\it relative} error of chirp mass reconstruction:

\begin{equation}\label{eq:R}
R = \left (\frac{1}{4\pi}\oint \left (\frac{\delta \cal{M}}{\cal{M}} \right )^2 d\Omega  \right )^{-\frac{1}{2}} = \left \langle \frac{\delta \cal{M}}{\cal{M}} \right \rangle,
\end{equation}

so that maximization of $R$ should lead to the minimum relative uncertainty of the reconstructed chirp mass averaged to the whole sky.

\subsection{Why we should not care about the directional sensitivity of the network}
\label{subsect:X}

One way to characterize the directional sensitivity of a telescope network is to calculate the mean square of its combined antenna factors for the $+$ and $\times$ polarizations toward the different sky directions. This measure will be independent from the polarization angle for any geometry of the interferometer arms, and thus independent from the orientations of the individual telescopes. By calculating the root-sum-square average of this quantity for the whole sky, we get a metric that characterizes the average directional sensitivity of the network for the whole sky, a.k.a. the average {\it reach} of the network \cite{BFS11}. For a network of $N$ $\Delta$-telescopes consisting of $3N$ interferometers, the metric is defined as:

\begin{equation}\label{eq:X}
X=\sqrt{\frac{1}{4\pi}\oint \frac{F^2_{+}\left ( \Phi,\lambda \right )+F^2_{\times}\left ( \Phi,\lambda \right )}{3N} d\Omega },
\end{equation}

where $F_+=F^\mathrm{network}_+$ and $F_{\times}=F^\mathrm{network}_{\times}$, both given in the DPF. Due to Eq.\ref{eq:Fnetwork}, the integral in Eq.\ref{eq:X} can be split to $3N$ integrals, each corresponding to a single interferometer in the network, and the integrals can be calculated individually. The value of $X$ will therefore never depend on the configuration of how the $N$ telescopes in the network are arranged, even if the interferometers have a different geometry from the interferometers in a $\Delta$-telescope. Even though by using a different network configuration we get a different {\it shape} for the combined antenna pattern, the all-sky integral of that surface will be the same for all configurations. This practically means, that by choosing a different configuration, we might lose sensitivity toward certain directions, but at the same time we increase the sensitivity of the network toward other directions. The net effect is that the $N$-telescope network will be able to cover the same volume of the local Universe, which is good enough in our case when we assume that the accessible Universe is homogeneous and isotropic. Thus, it is unnecessary to include the $X$ metric in our optimization procedure, and we can use only the other three metrics that we defined, denoted as $I$, $D$, and $R$. Nevertheless, for future publications when targeted science is considered, a metric characterizing the directional sensitivity of the network, as well as the time evolution of cosmic reach will be important.

\subsection{The combined metric}
\label{subsect:C}

The three metrics we described so far, $I$, $D$, and $R$, together define a three dimensional metric space. For each $N$-telescope configurations, we can calculate the corresponding metric values, and identify a spatial point in this space using the metric values as coordinates. The goal of the optimization process for the individual metrics is to find the configuration of $N$ telescopes that corresponds to the highest values of $I$, $D$, and $R$, respectively, from a number of Monte-Carlo trials. This is equivalent to the goal of finding the most distant data points in the metric space along the three dimensions, separately. From now on we denote these highest values of the individual metrics that we actually find in our sample of different $N$-telescope configurations as $I_\mathrm{max}$, $D_\mathrm{max}$, and $R_\mathrm{max}$, respectively.

The optimal configuration of an $N$-telescope network in a combined metric would be the one for which all the corresponding metric values are maximal within the sample. Unfortunately, due to the competing nature of the three metrics, there is no such configuration, thus, we need to construct a combined metric for such a generalized optimization process, where the three individual metrics are combined with specific weights.

In our analysis, we did not want to favor any of the three metrics specifically, therefore we normalized the metric values with $I_\mathrm{max}$, $D_\mathrm{max}$, and $R_\mathrm{max}$, respectively. This makes all the metric values dimensionless, and somewhat comparable with each other in nature. Using this, the combined metric we defined was:

\begin{equation}\label{eq:C}
C = \sqrt{\left (\frac{I}{I_\mathrm{max}}  \right )^2 + \left (\frac{D}{D_\mathrm{max}}  \right )^2 + \left (\frac{R}{R_\mathrm{max}}  \right )^2}.
\end{equation}

Specific astronomic searches may require different weighting factors, however, as a general approach to the problem, we used equal weights.

\begin{table}
\caption{\label{Table:metrics}The FoMs and their corresponding metrics used in the optimization of $N$-telescope configurations. For detailed descriptions, see section \ref{section_maths}.}
\footnotesize\rm
\begin{tabular*}{\textwidth}{@{}l*{15}{@{\extracolsep{0pt plus12pt}}l}}
\br
FoM&Metric&Metric description\\
\mr
Polarization reconstruction&$I$&$\left (\frac{1}{4\pi}\oint \left | F^\mathrm{network}_+\left ( \Phi ,\lambda  \right ) - F^\mathrm{network}_{\times}\left ( \Phi ,\lambda  \right ) \right |^2 d\Omega  \right )^{-\frac{1}{2}}$\\
Source localization (N=2)&$D$&Square of the distance between the telescopes\\
Source localization (N=3)&$D$&Area of the triangle formed by the telescopes\\
Source localization (N=5)&$D$&Area of the $N=3$ sub-network with highest $D$\\
Source parameter reconstruction&$R$&$\left (\frac{1}{4\pi}\oint \left (\frac{\delta \cal{M}}{\cal{M}} \right )^2 d\Omega  \right )^{-\frac{1}{2}} = \left \langle \frac{\delta \cal{M}}{\cal{M}} \right \rangle$\\
Combined metric&$C$&$\sqrt{\left (\frac{I}{I_\mathrm{max}}  \right )^2 + \left (\frac{D}{D_\mathrm{max}}  \right )^2 + \left (\frac{R}{R_\mathrm{max}}  \right )^2}$\\
\br
\end{tabular*}
\end{table}

%% file: opt_position.tex
We carried out an optimization procedure in three example cases where the location of the first $\Delta$-telescope was predetermined and a second telescope ranges over the set of $1524$ acceptable sites (see section \ref{section_map}). In the three example cases the location of the first telescope site was chosen to be $[\Phi_\mathrm{A},\lambda_\mathrm{A}]=[48.5^{\circ},18.7^{\circ}]$ in Europe (Example A), $[\Phi_\mathrm{B},\lambda_\mathrm{B}]=[38.9^{\circ},-98.4^{\circ}]$ in North America (Example B), and $[\Phi_\mathrm{C},\lambda_\mathrm{C}]=[19.6^{\circ},78.0^{\circ}]$ in India (Example C).

\begin{figure}
    \centering
\begin{tabular}{cc}
    \subfloat[]{\includegraphics[width=80mm]{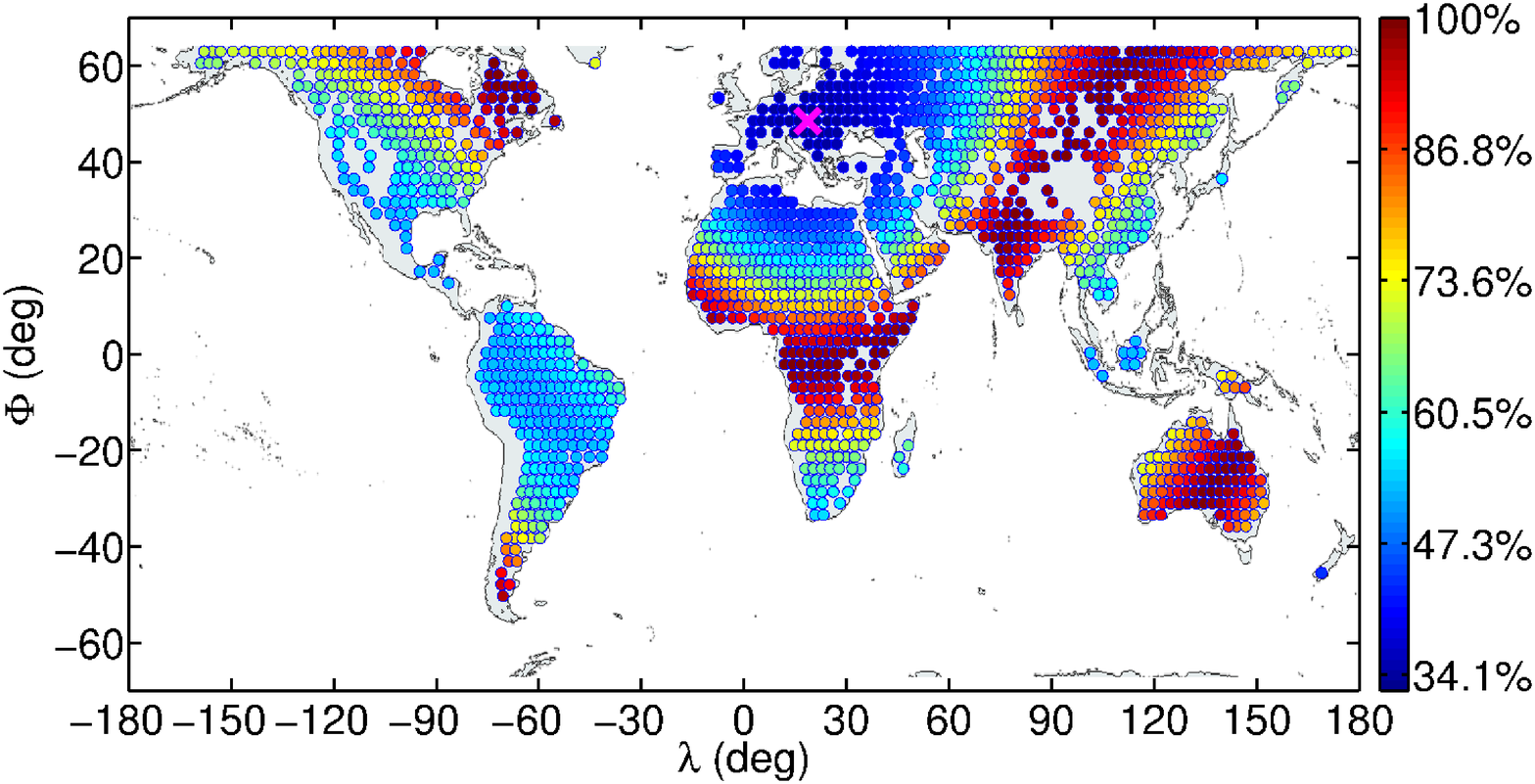}} & \hspace{-6mm} \subfloat[]{\includegraphics[width=80mm]{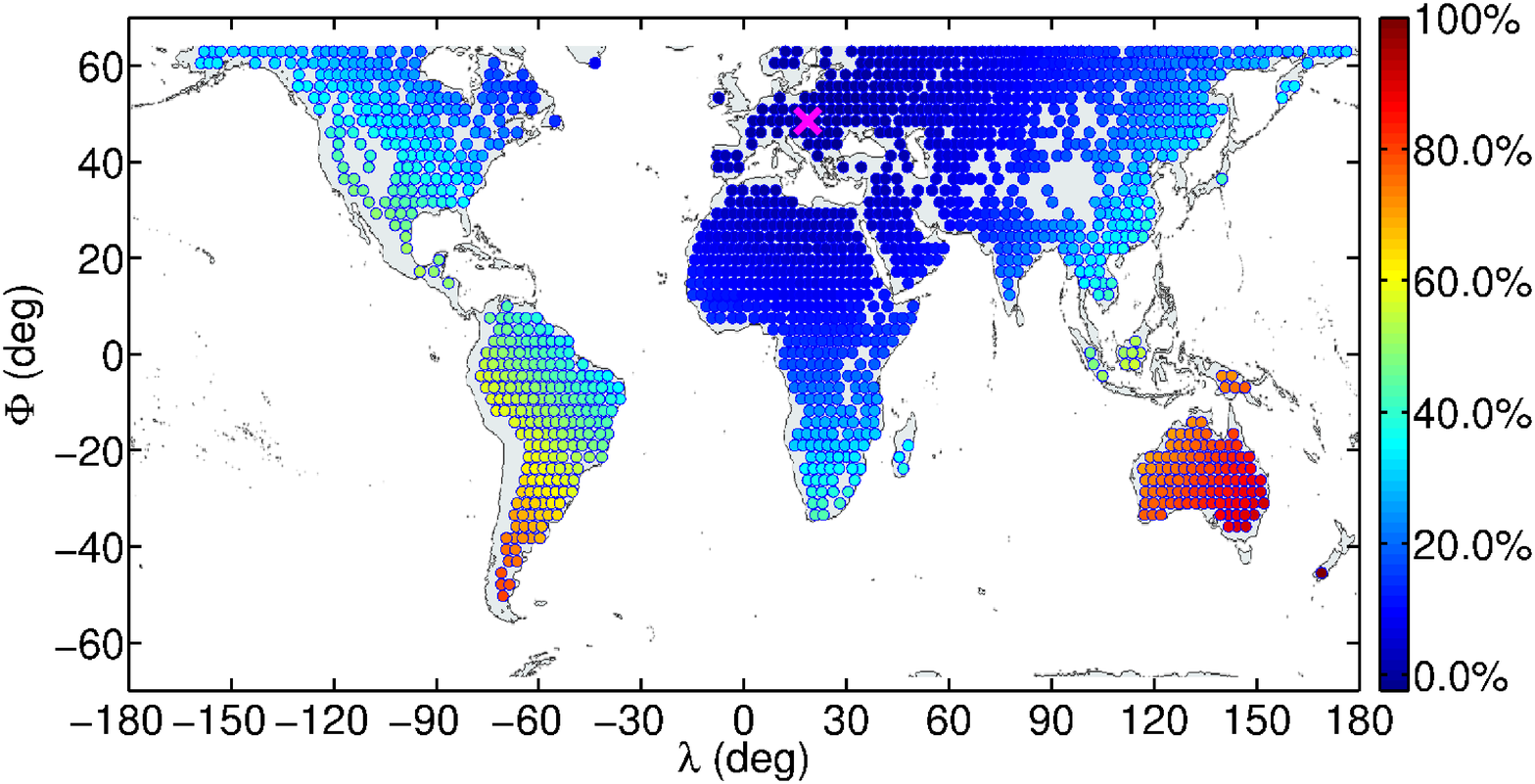}} \\
\end{tabular}
    \subfloat[]{\includegraphics[width=80mm]{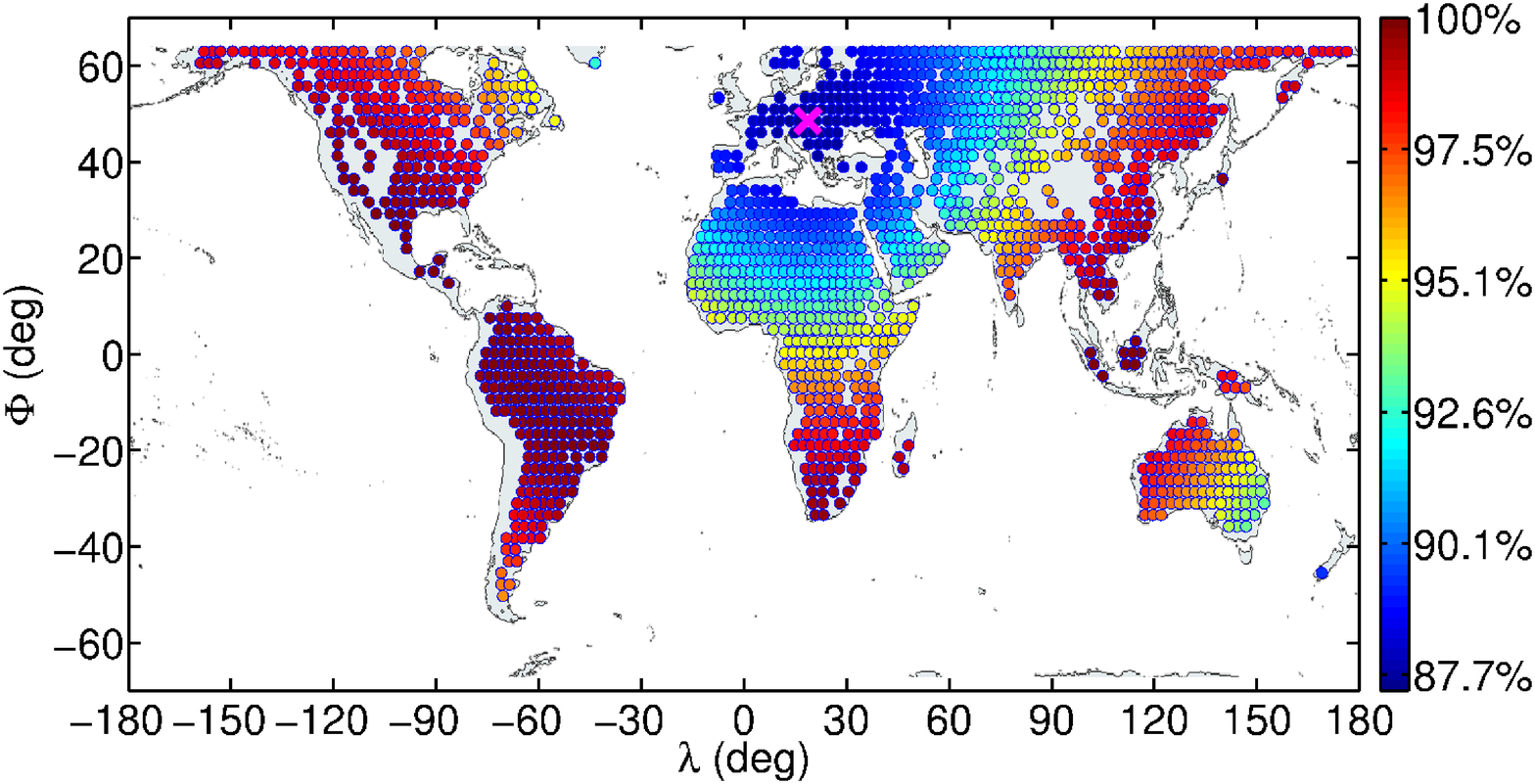}}
\caption{The colormaps of normalized metric values that we get for a network of two identical $\Delta$-telescopes in the case when the first telescope is placed to $[\Phi_\mathrm{A},\lambda_\mathrm{A}]=[48.5^{\circ},18.7^{\circ}]$ in Europe, and the location of the second telescope is chosen from the set of acceptable sites described in section \ref{section_map}. Each circle in the maps corresponds to one of the $1524$ site locations, while their colors represent the $I/I_\mathrm{max}$ (a), $D/D_\mathrm{max}$ (b), or $R/R_\mathrm{max}$ (c) values, given in percentages, where the ''max'' indices correspond to the highest values of the three metrics within the samples of $1524$ elements.
}\label{fig:2detFixedE}
\end{figure}

\begin{figure}
    \centering
\begin{tabular}{cc}
    \subfloat[]{\includegraphics[width=80mm]{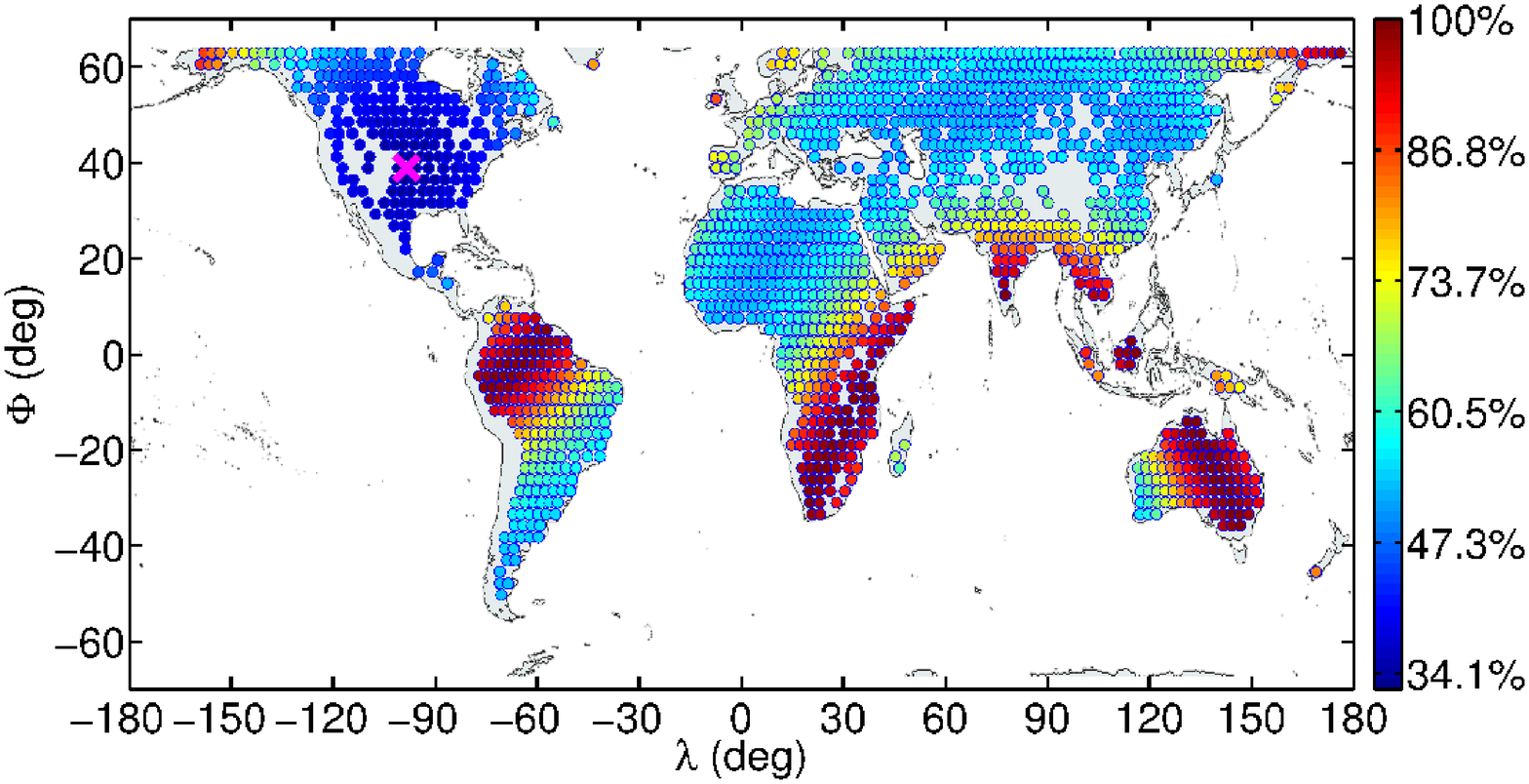}} & \hspace{-6mm} \subfloat[]{\includegraphics[width=80mm]{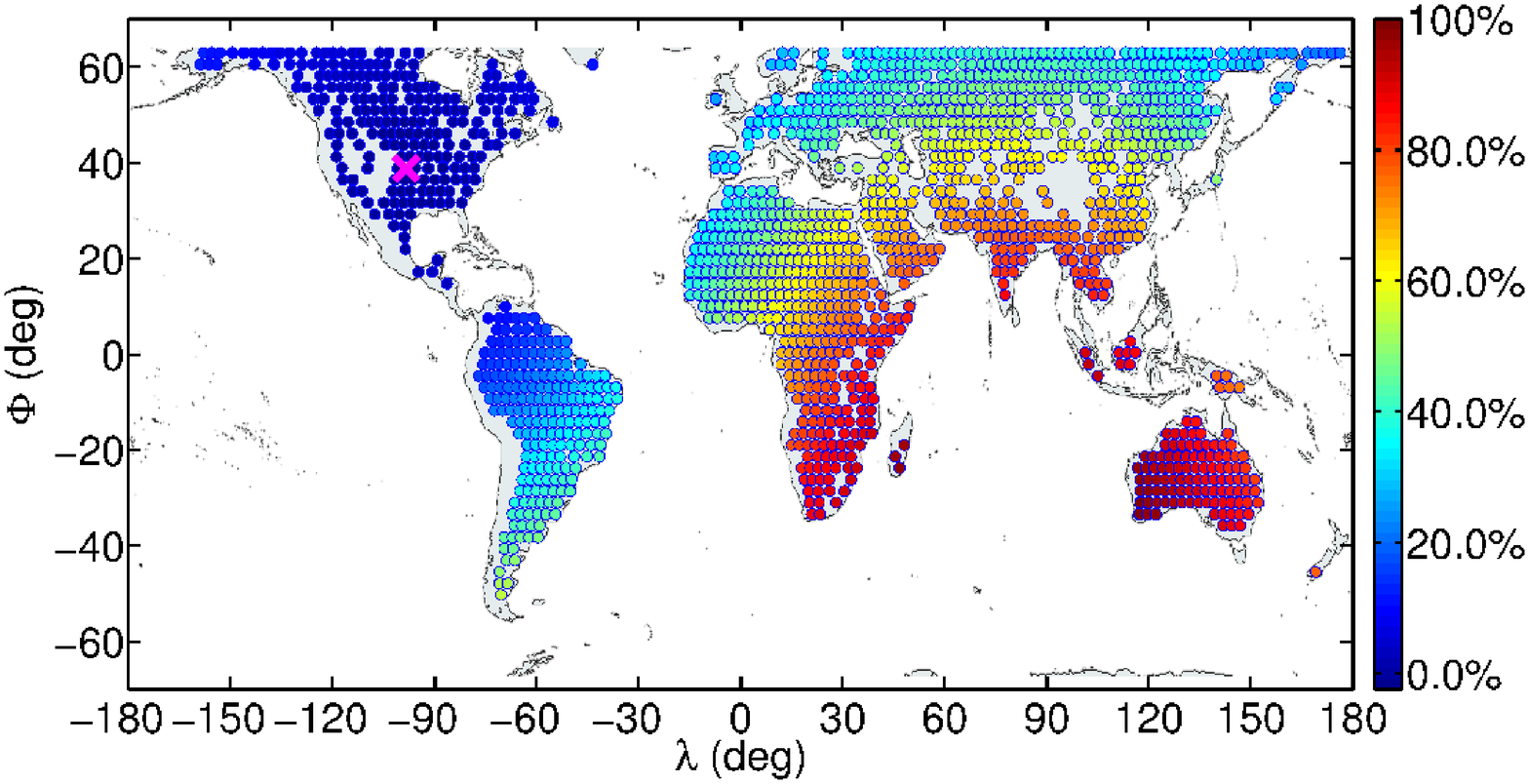}} \\
\end{tabular}
    \subfloat[]{\includegraphics[width=80mm]{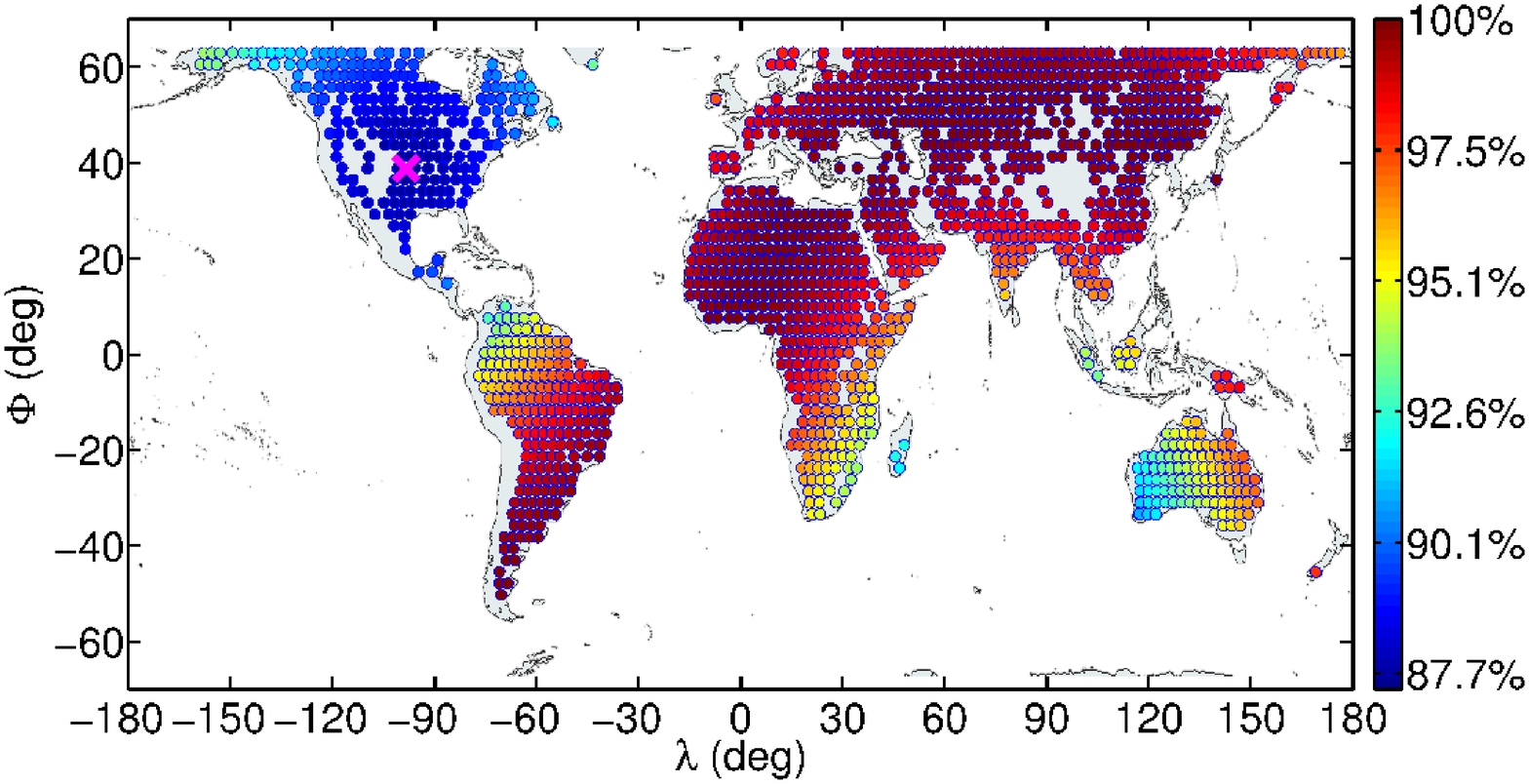}}
\caption{The colormaps of normalized metric values that we get for a network of two identical $\Delta$-telescopes in the case when the first telescope is placed to $[\Phi_\mathrm{B},\lambda_\mathrm{B}]=[38.9^{\circ},-98.4^{\circ}]$ in North America, and the location of the second telescope is chosen from the set of acceptable sites described in section \ref{section_map}. Each circle in the maps corresponds to one of the $1524$ site locations, while their colors represent the $I/I_\mathrm{max}$ (a), $D/D_\mathrm{max}$ (b), or $R/R_\mathrm{max}$ (c) values, given in percentages, where the ''max'' indices correspond to the highest values of the three metrics within the samples of $1524$ elements.
}\label{fig:2detFixedU}
\end{figure}

\begin{figure}
    \centering
\begin{tabular}{cc}
    \subfloat[]{\includegraphics[width=80mm]{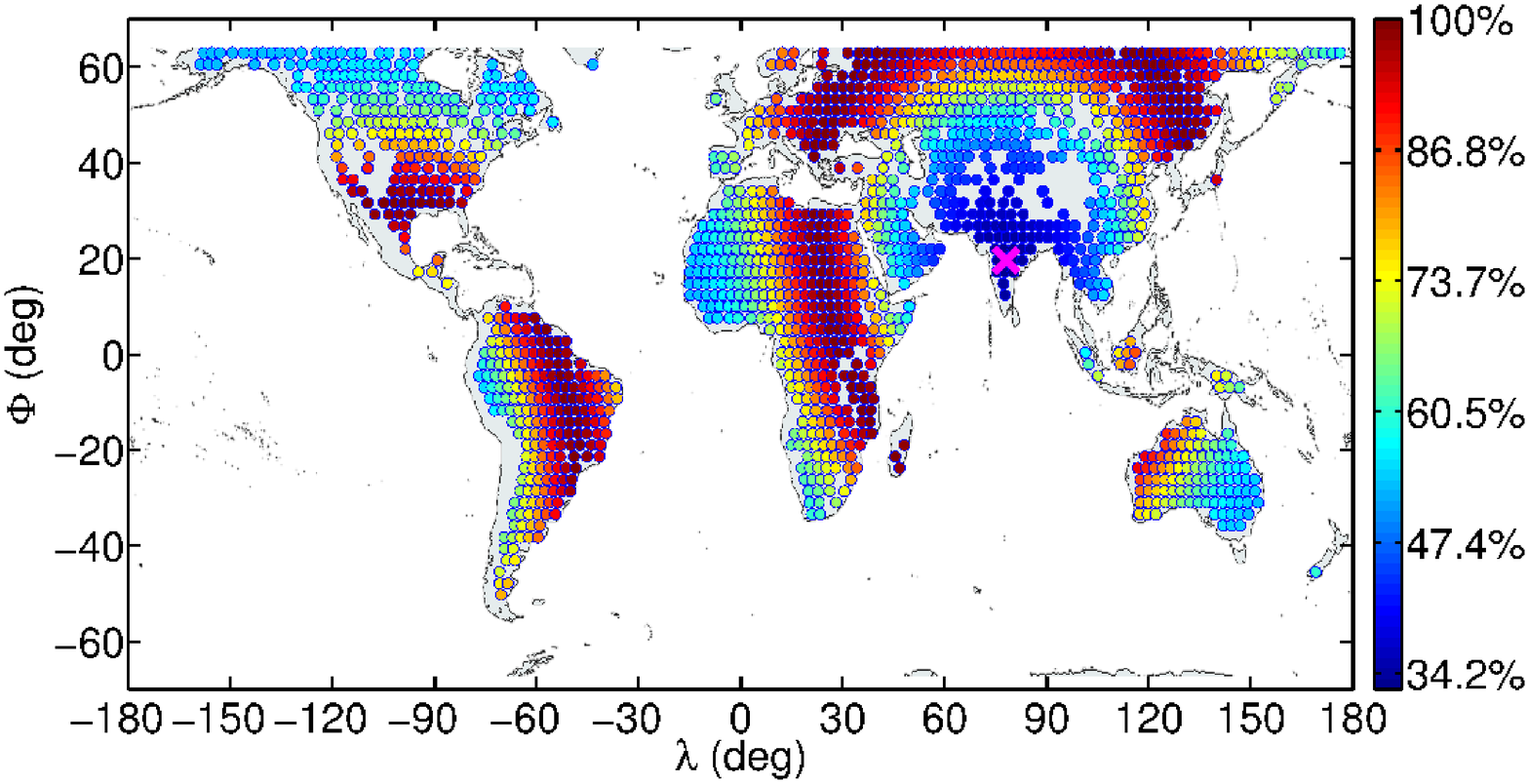}} & \hspace{-6mm} \subfloat[]{\includegraphics[width=80mm]{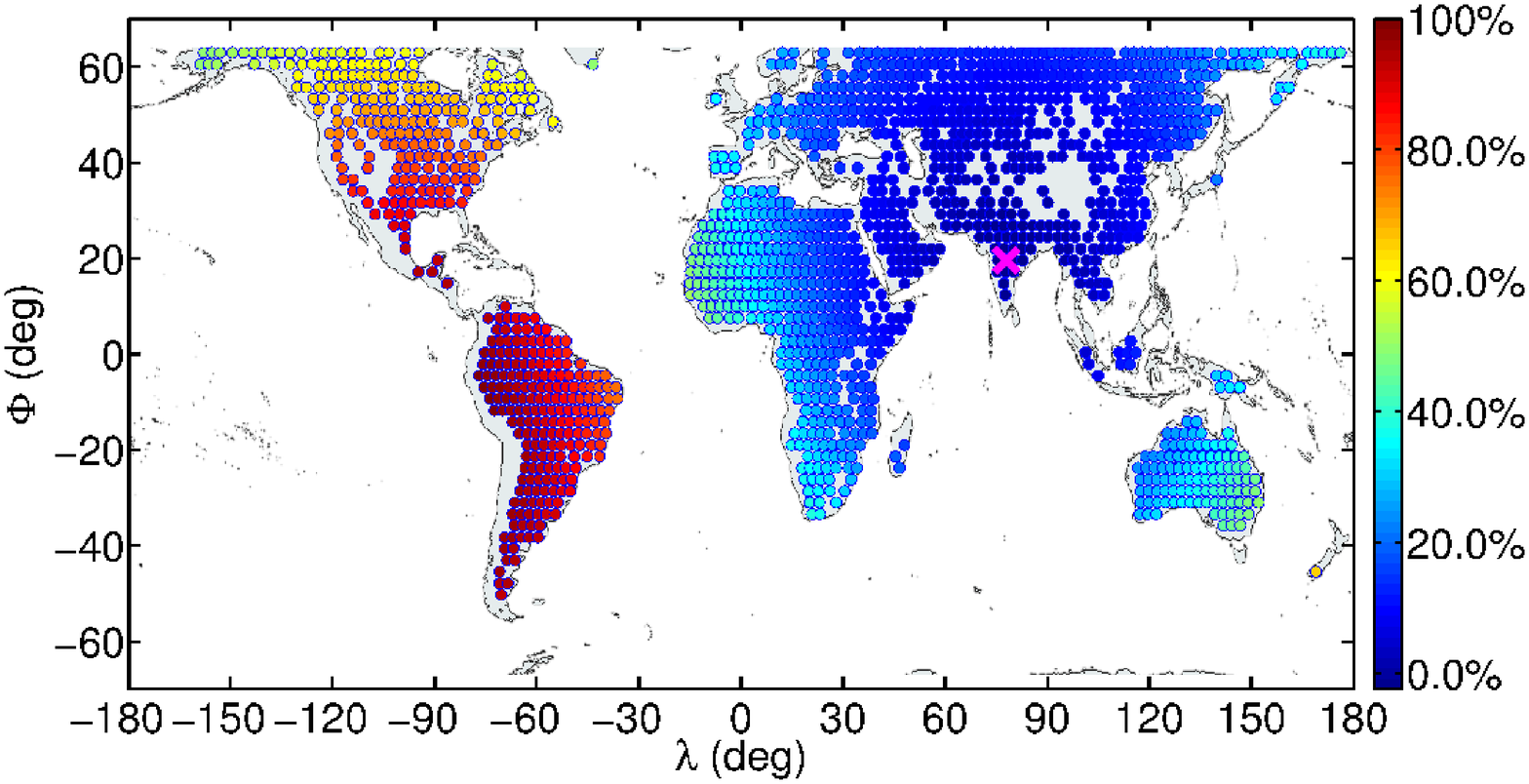}} \\
\end{tabular}
    \subfloat[]{\includegraphics[width=80mm]{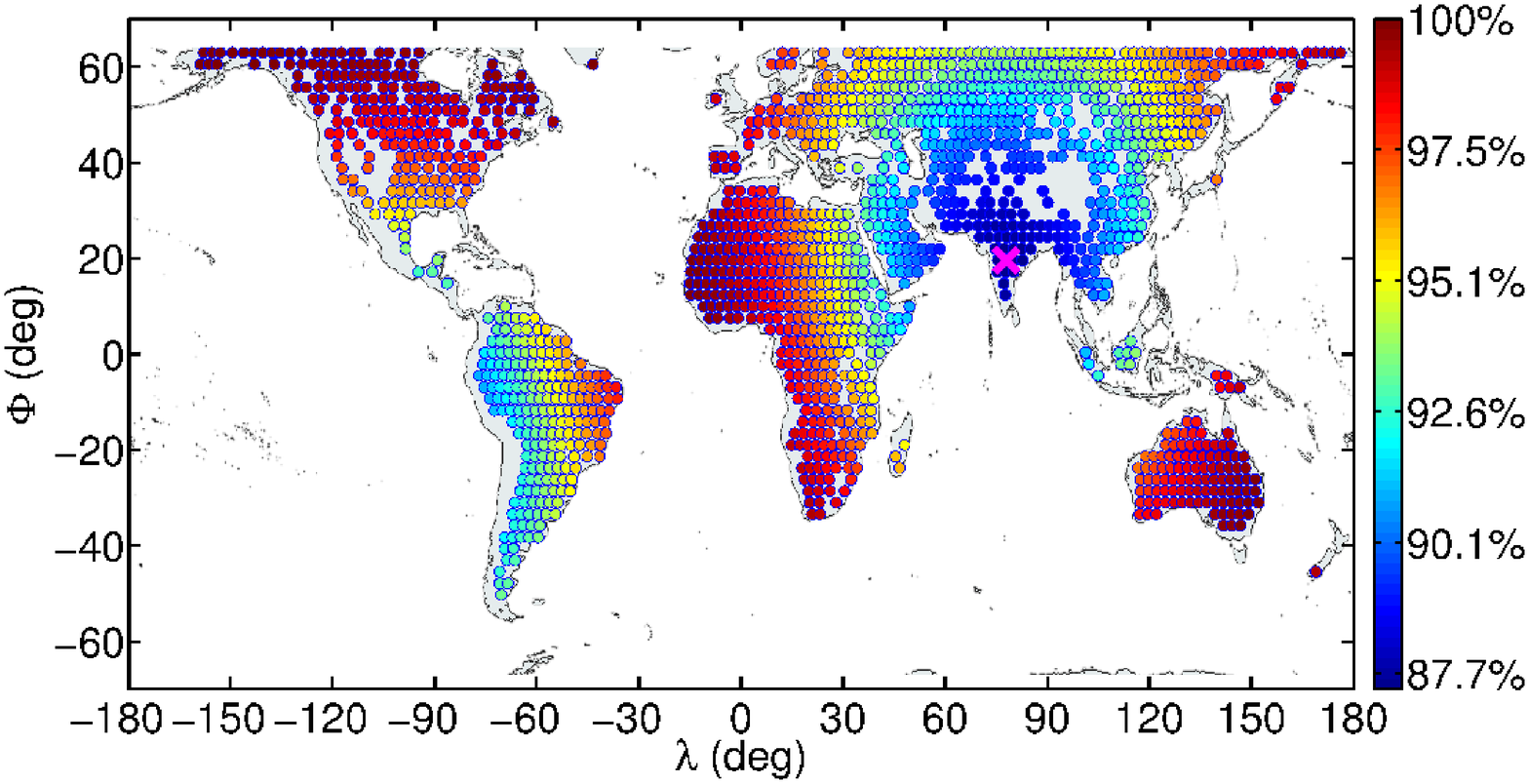}}
\caption{The colormaps of normalized metric values that we get for a network of two identical $\Delta$-telescopes in the case when the first telescope is placed to $[\Phi_\mathrm{C},\lambda_\mathrm{C}]=[19.6^{\circ},78.0^{\circ}]$ in India, and the location of the second telescope is chosen from the set of acceptable sites described in section \ref{section_map}. Each circle in the maps corresponds to one of the $1524$ site locations, while their colors represent the $I/I_\mathrm{max}$ (a), $D/D_\mathrm{max}$ (b), or $R/R_\mathrm{max}$ (c) values, given in percentages, where the ''max'' indices correspond to the highest values of the three metrics within the samples of $1524$ elements.
}\label{fig:2detFixedI}
\end{figure}

We then chose a location for the second $\Delta$-telescope from the set all $1524$ acceptable sites, and calculated the values of the $I$, $D$, and $R$ metrics (see section \ref{section_maths}) for the corresponding $2$-telescope network. The resulting set of $I$, $D$, and $R$ values were then normalized with the highest value of the three metrics among the sample of $1524$ elements, denoted as $I_\mathrm{max}$, $D_\mathrm{max}$, and $R_\mathrm{max}$, respectively. The $I/I_\mathrm{max}$, $D/D_\mathrm{max}$, and $R/R_\mathrm{max}$ ratios (given in percentages) are shown in figure \ref{fig:2detFixedE} for Example A (Europe), in figure \ref{fig:2detFixedU} for Example B (USA), and in figure \ref{fig:2detFixedI} for Example C (India).

\begin{figure}
    \centering
    \includegraphics[width=120mm]{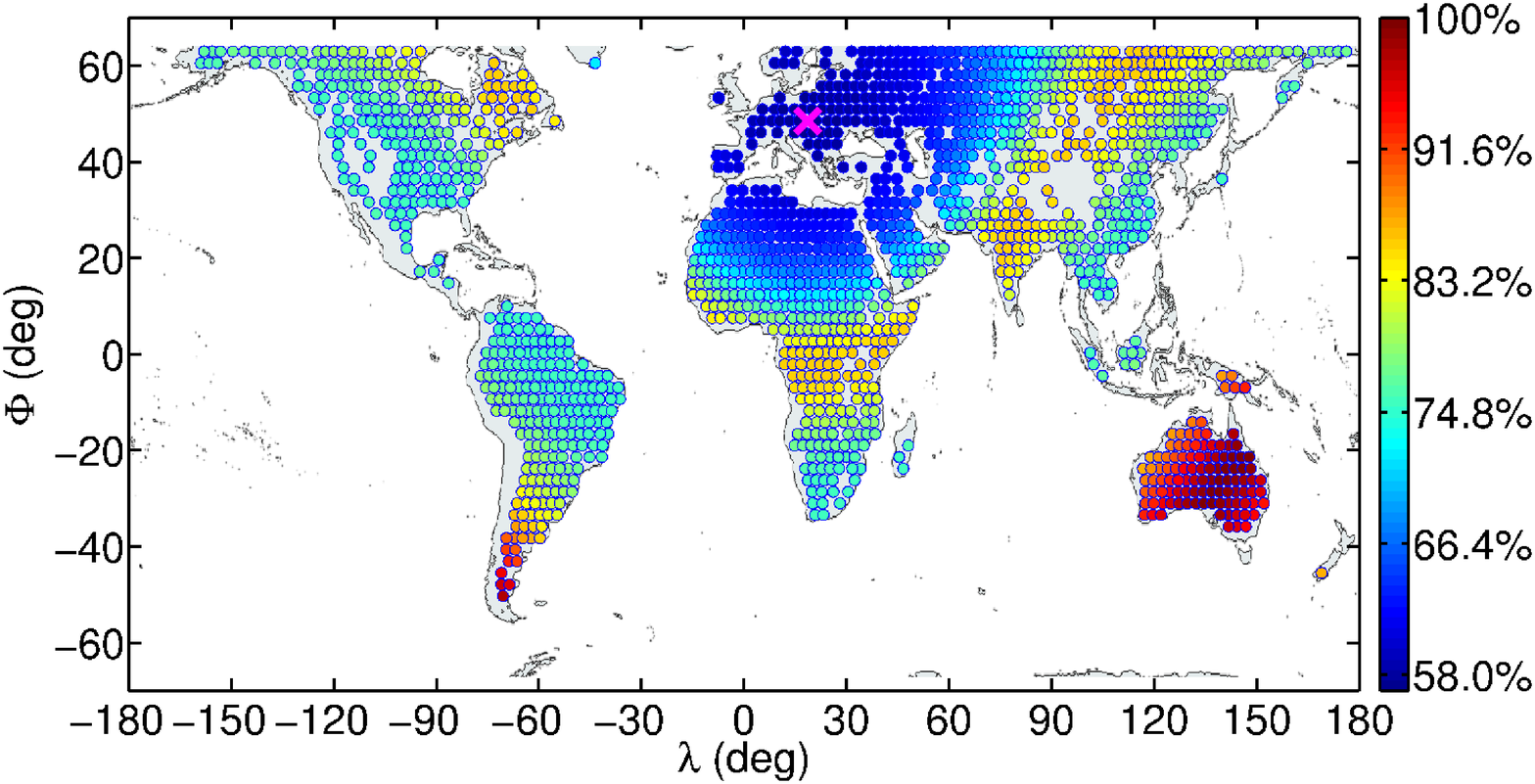}
\caption{The colormap of $C/C_\mathrm{max}$ values for $1524$ different network configurations of two identical $\Delta$-telescopes, where $C_\mathrm{max}$ is the highest value of the combined metric within the sample of $1524$ elements. The location of the first telescope was chosen to be at $[\Phi,\lambda]=[48.5^{\circ},18.7^{\circ}]$ (marked by $\times$), and the location of the second telescope was chosen from the set of acceptable sites described in section \ref{section_map}.}\label{fig:2detFixedCE}
\end{figure}

\begin{figure}
    \centering
    \includegraphics[width=120mm]{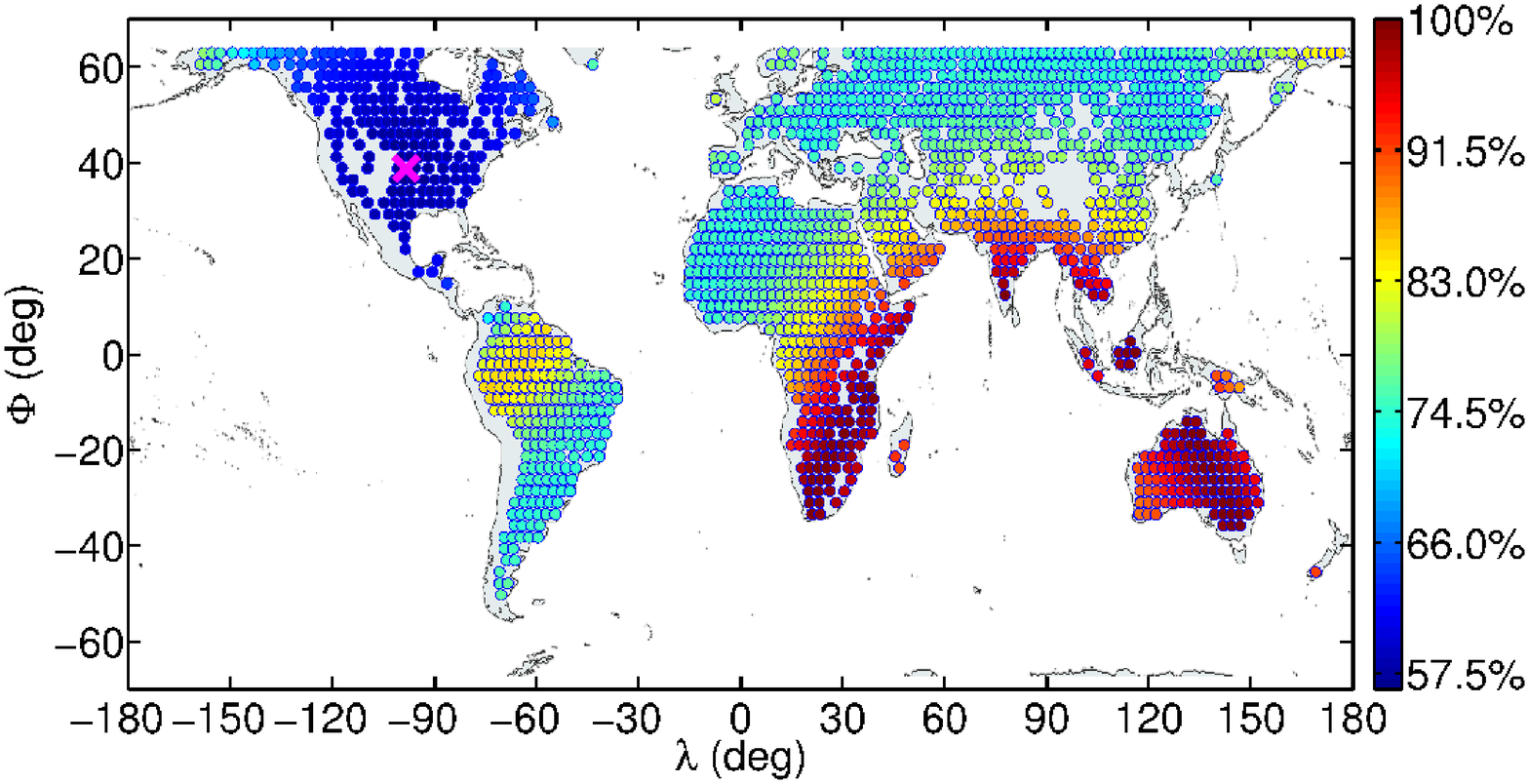}
\caption{The colormap of $C/C_\mathrm{max}$ values for $1524$ different network configurations of two identical $\Delta$-telescopes, where $C_\mathrm{max}$ is the highest value of the combined metric within the sample of $1524$ elements. The location of the first telescope was chosen to be at $[\Phi,\lambda]=[38.9^{\circ},-98.4^{\circ}]$ (marked by $\times$), and the location of the second telescope was chosen from the set of acceptable sites described in section \ref{section_map}.}\label{fig:2detFixedCU}
\end{figure}

\begin{figure}
    \centering
    \includegraphics[width=120mm]{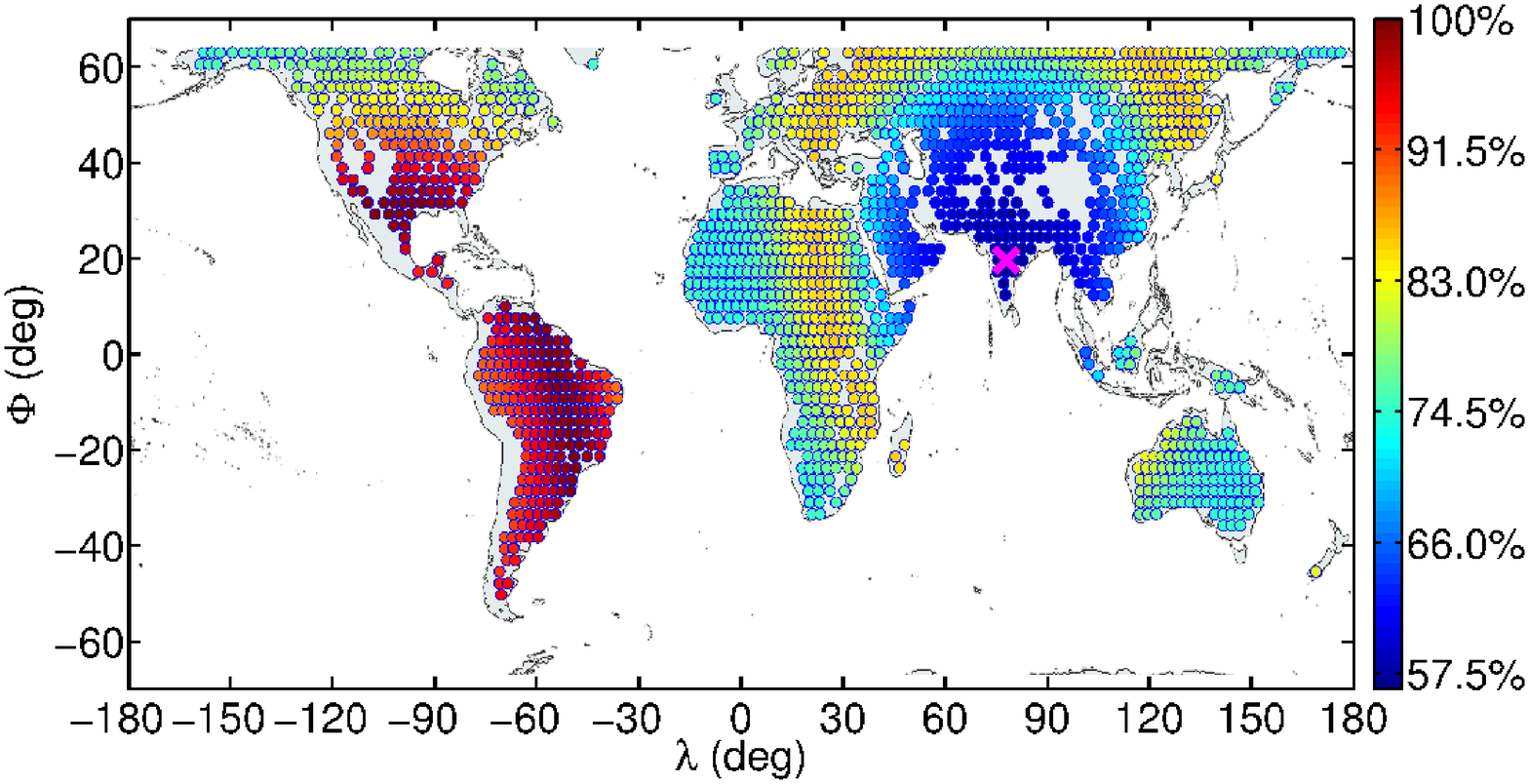}
\caption{The colormap of $C/C_\mathrm{max}$ values for $1524$ different network configurations of two identical $\Delta$-telescopes, where $C_\mathrm{max}$ is the highest value of the combined metric within the sample of $1524$ elements. The location of the first telescope was chosen to be at $[\Phi,\lambda]=[19.6^{\circ},78.0^{\circ}]$ (marked by $\times$), and the location of the second telescope was chosen from the set of acceptable sites described in section \ref{section_map}.}\label{fig:2detFixedCI}
\end{figure}

Using the known $\{I,D,R\}/\{I,D,R\}_\mathrm{max}$ ratios, we calculated the $C$ metric (see section \ref{subsect:C}) for all the $1524$ different $2$-telescope configurations, and for all three example cases. We again normalized the resulting $C$ values with the highest value in the sample, $C_\mathrm{max}$, and visualized the $C/C_\mathrm{max}$ values (given in percentages) for the different configurations in figure \ref{fig:2detFixedCE} for Example A (Europe), in figure \ref{fig:2detFixedCU} for Example B (USA), and in figure \ref{fig:2detFixedCI} for Example C (India).

If we choose a specific site for the first $\Delta$-telescope in the network, and allow any given percentage of loss in the value of a chosen metric (compared to $\{I/D/R/C\}_\mathrm{max}$), we can count the number of sites that satisfy this tolerance criterion, and consider each of them as being an optimal site for a second $\Delta$-telescope. By doing so, we can associate the number of such sites to all the $1525$ acceptable sites. Using this general method, we can choose the site for the first $\Delta$-telescope that provides the most options for optimal site selection when extending the network with a second instrument.

\begin{figure}
    \centering
    \includegraphics[width=120mm]{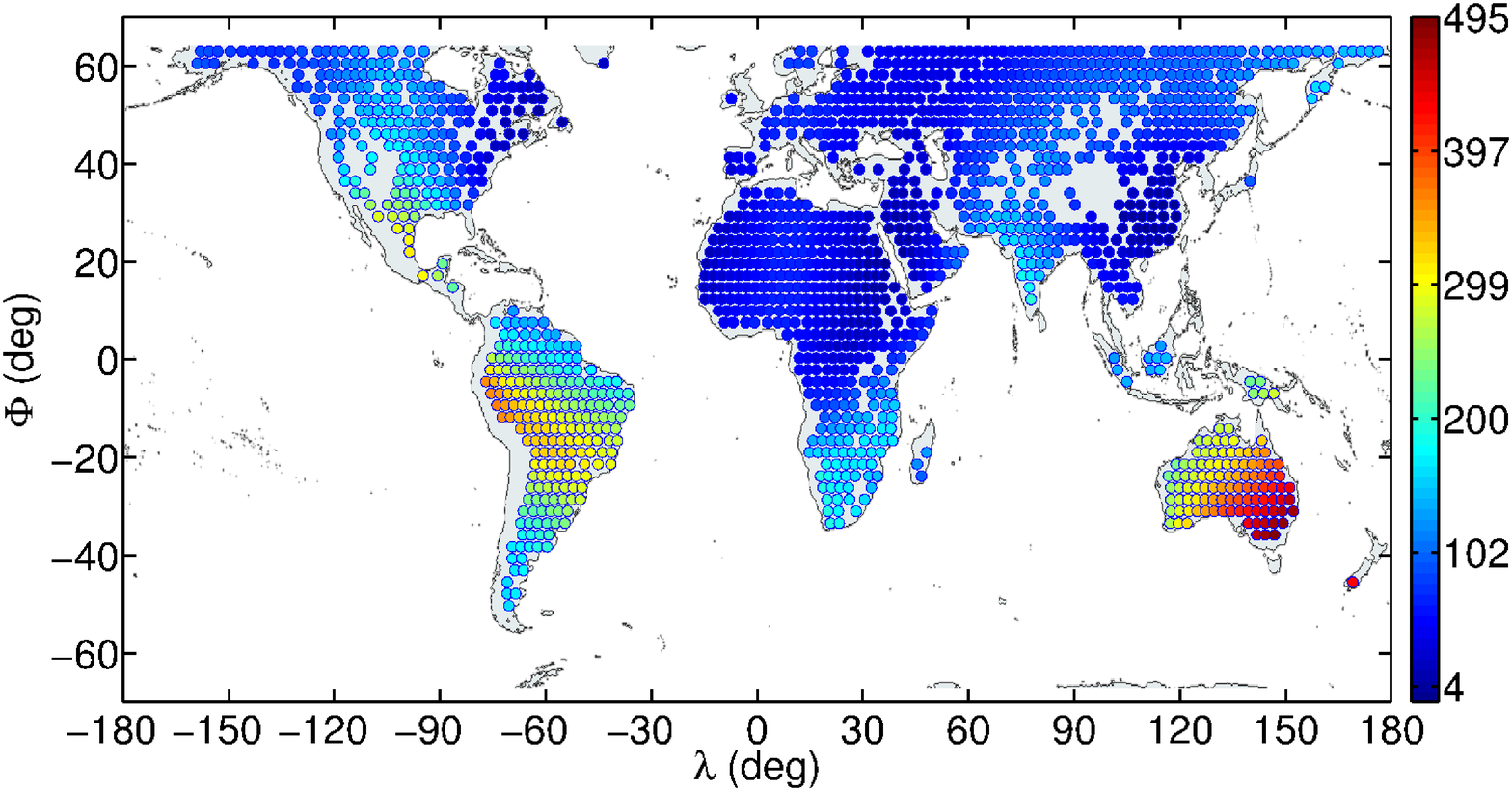}
\caption{This colormap shows the number of sites (and thus the number of corresponding optimal $2$-telescope networks) satisfying our example tolerance criterion on the $C$ metric, assuming that the first $\Delta$-telescope is placed to the different acceptable sites represented by the colored circles (see section \ref{section_map}). The highest numbers (up to $495$) of optimal sites for a second instrument (i.e. sites that satisfy the tolerance criterion) are associated to locations in Australia. Note, that the colormap strongly depends on the specific metric and tolerance criterion we apply, however Australia being the optimal region for the first $\Delta$-telescope is very robust to the tolerance criterion given that the combined ($C$) metric is applied.}\label{fig:OptNumMap}
\end{figure}

As an example, we chose a tolerance criterion on the combined ($C$) metric, and allowed a maximum of $-5\%$ loss in its value compared to $C_\mathrm{max}$. Using this tolerance criterion, we found that the optimal site locations for a second telescope always form a $\pm 10^{\circ}$ annulus at an angular distance of $\sim 130^{\circ}$ from the location of the first telescope. The number of sites satisfying the tolerance criterion for each of the $1525$ acceptable sites (and thus the number of corresponding optimal $2$-telescope networks) is depicted in figure \ref{fig:OptNumMap}. As figure \ref{fig:OptNumMap} shows, placing the first telescope in Australia provides the most options for optimal site selection when extending the network with a second instrument. Note, that even though this example gives a general method for choosing an optimal site for the first and second $\Delta$-telescope in the network, the resulting optimal sites will strongly depend on the metric we use in this optimization. However, when choosing the $C$ metric, the optimal locations for the first $\Delta$-telescope will be in Australia for a wide range of allowed losses in $C/C_\mathrm{max}$. Thus the result of Australia being the optimal region for the first $\Delta$-telescope is very robust to different tolerance criteria given that the combined ($C$) metric is applied.

\begin{figure}
    \centering
    \includegraphics[width=120mm]{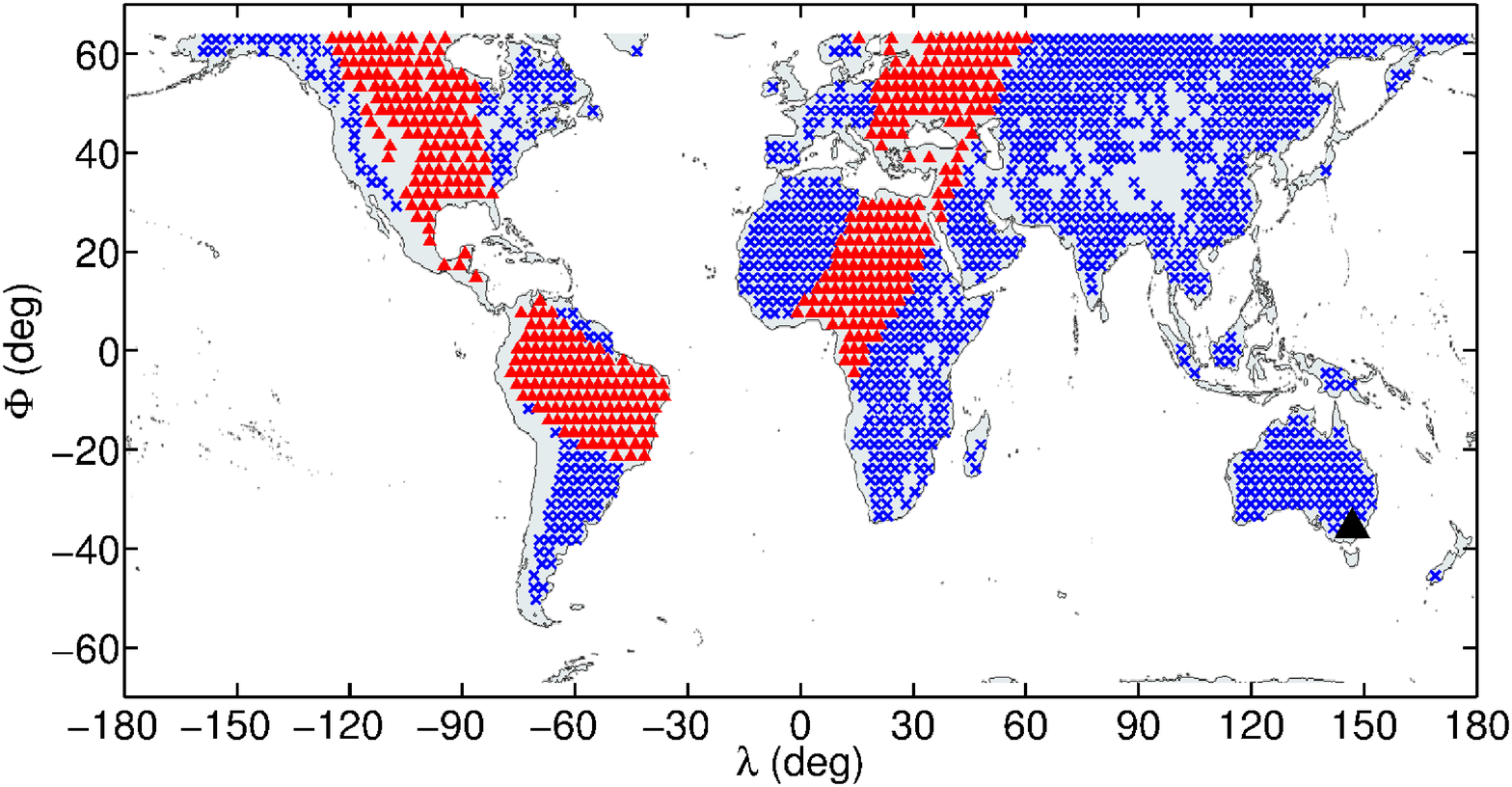}
\caption{This map shows the annulus with the highest number of optimal site locations for a second $\Delta$-telescope, based on the example tolerance criterion we applied on the $C$ metric. The annulus is formed by the set of optimal sites (represented by small triangles) satisfying the tolerance criterion when the first telescope is placed to $[\Phi,\lambda]=[-35.8^{\circ},146.9^{\circ}]$ in Australia (represented by a big black triangle). The set of sites not satisfying the criterion are represented by $\times$ symbols in the map. If the first $\Delta$-telescope is placed to $[\Phi,\lambda]=[-35.8^{\circ},146.9^{\circ}]$ in Australia, we suggest that the location of the second instrument should be chosen from the elements of the annulus.}\label{fig:Australia}
\end{figure}

Figure \ref{fig:Australia} shows the $\pm 10^{\circ}$ annulus of optimal sites (corresponding to an allowed maximum of $-5\%$ loss in $C/C_\mathrm{max}$) in the case when the first $\Delta$-telescope is placed to a geographical position of $[\Phi,\lambda]=[-35.8^{\circ},146.9^{\circ}]$ in Australia. We suggest that the site location for the second telescope should be chosen from within this annulus if the first $\Delta$-telescope is placed to $[\Phi,\lambda]=[-35.8^{\circ},146.9^{\circ}]$. Note, that placing the first $\Delta$-telescope of the network to any of the acceptable sites within the annulus would result with $[\Phi,\lambda]=[-35.8^{\circ},146.9^{\circ}]$ being in the annulus of optimal sites for the second $\Delta$-telescope.

By setting the location of the first $\Delta$-telescope in the network to be in Australia, and choosing a site for the second $\Delta$-telescope in the network from the corresponding $\pm 10^{\circ}$ annulus, we can rerun the optimization procedure for a $3$-telescope network with these two predetermined site locations, similarly to the previous cases of $2$-telescope networks with the first site locations predetermined.

\begin{figure}
    \centering
\begin{tabular}{cc}
    \subfloat[]{\includegraphics[width=80mm]{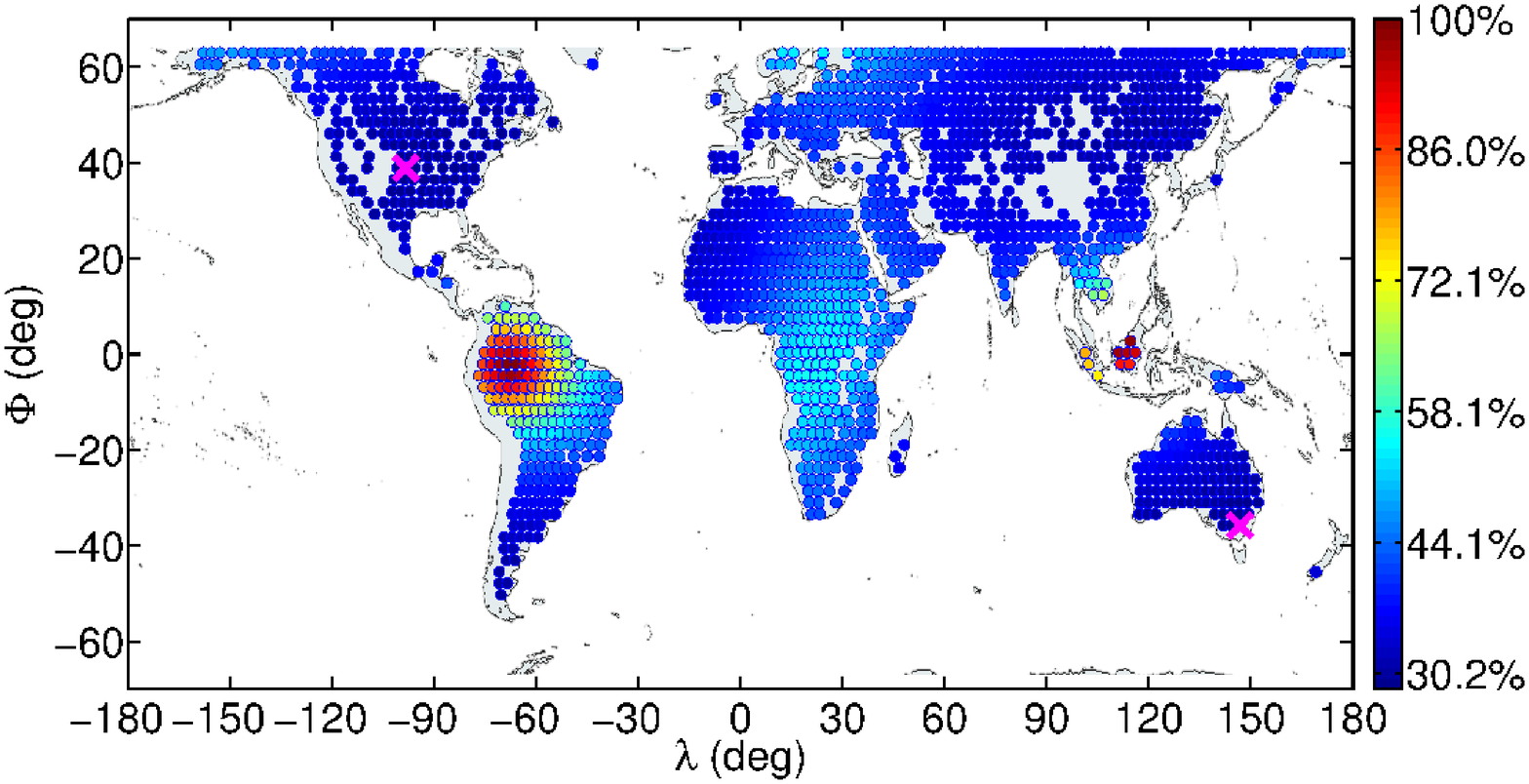}} & \hspace{-6mm} \subfloat[]{\includegraphics[width=80mm]{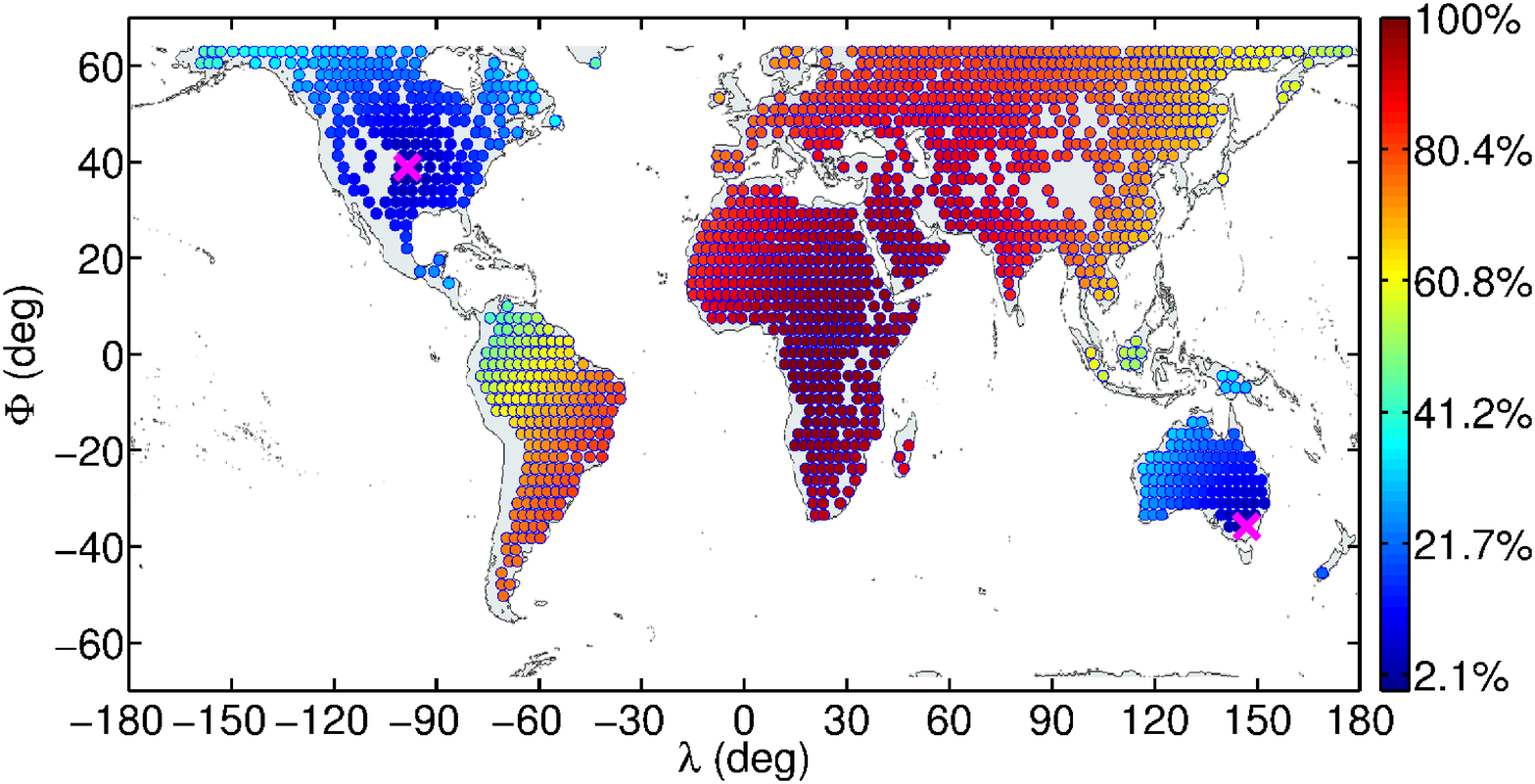}} \\
\end{tabular}
    \subfloat[]{\includegraphics[width=80mm]{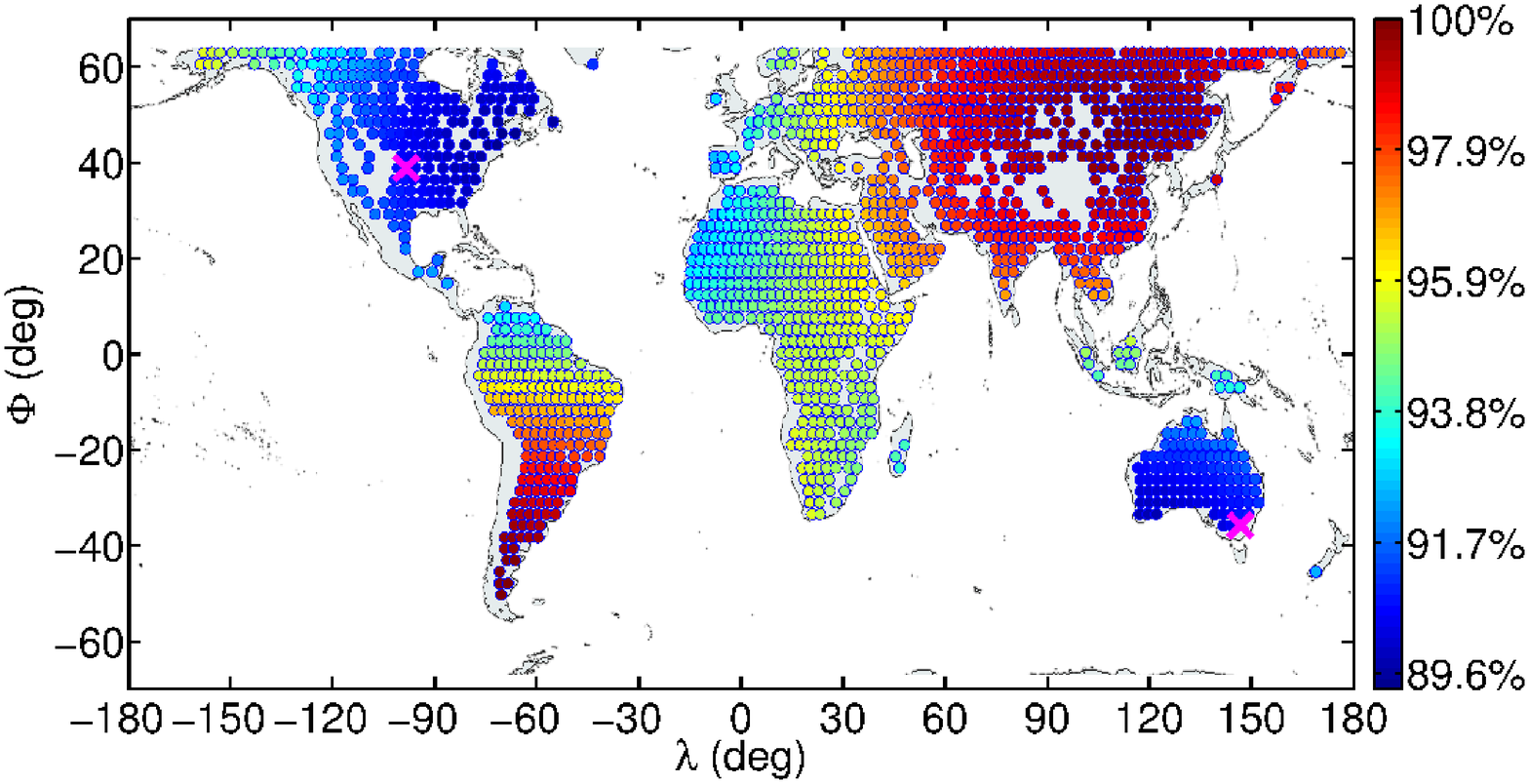}}
\caption{The colormaps of normalized metric values that we get for a network of three identical $\Delta$-telescopes in the case when one telescope is placed to $[\Phi_\mathrm{D1},\lambda_\mathrm{D1}]=[-35.8^{\circ},146.9^{\circ}]$ in Australia, and one to $[\Phi_\mathrm{D2},\lambda_\mathrm{D2}]=[38.9^{\circ},-98.4^{\circ}]$ in North America (marked by $\times$ symbols). The location of the third telescope is chosen from the set of acceptable sites described in section \ref{section_map}. Each circle in the maps corresponds to one of the $1523$ site locations, while their colors represent the $I/I_\mathrm{max}$ (a), $D/D_\mathrm{max}$ (b), or $R/R_\mathrm{max}$ (c) values, given in percentages, where the ''max'' indices correspond to the highest values of the three metrics within the samples of $1523$ elements.
}\label{fig:AustraliaUSA}
\end{figure}

\begin{figure}
    \centering
    \includegraphics[width=120mm]{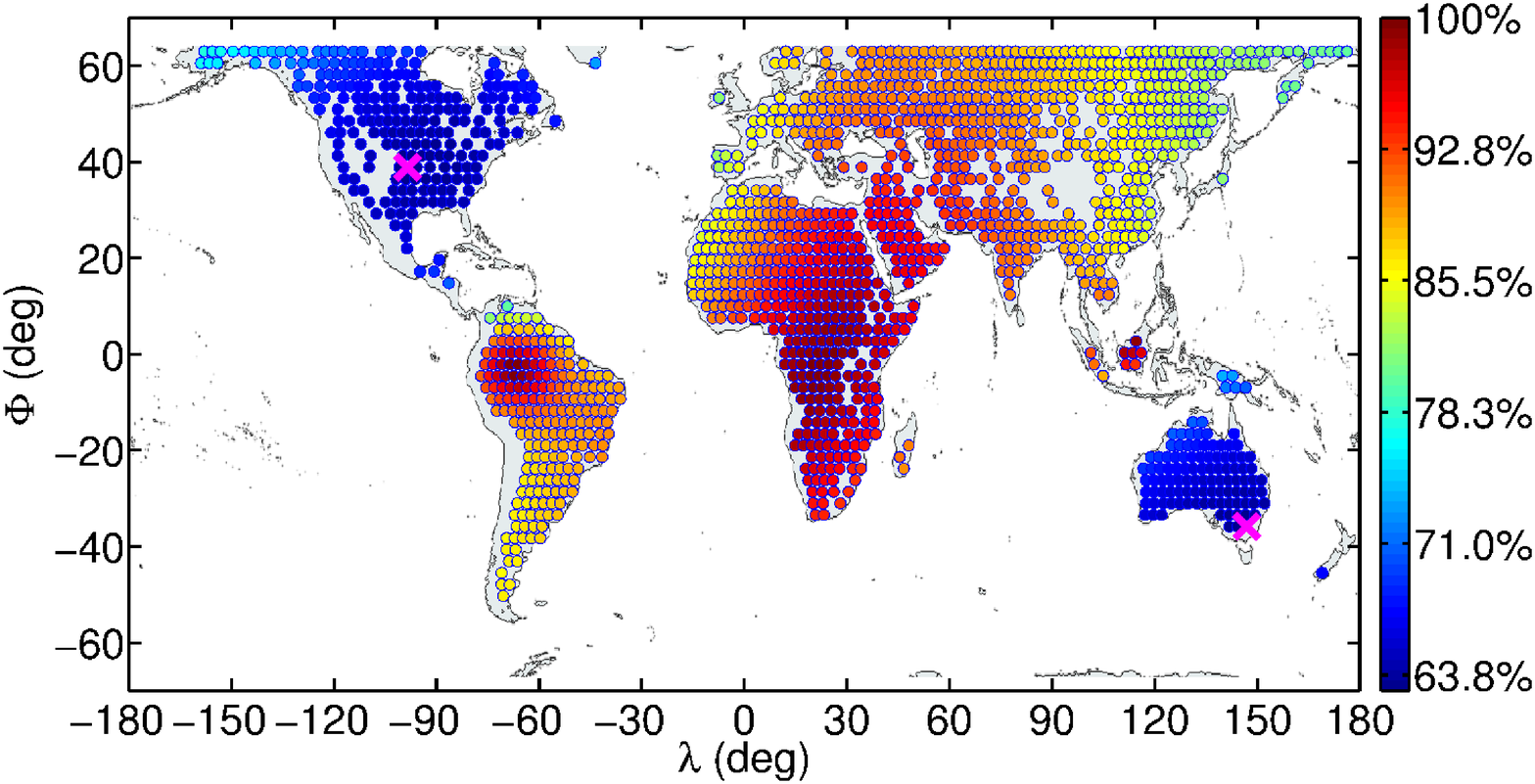}
\caption{The colormap of $C/C_\mathrm{max}$ values for $1523$ different network configurations of three identical $\Delta$-telescopes, where $C_\mathrm{max}$ is the highest value of the combined metric within the sample of $1523$ elements. The location of the first telescope in the network were chosen to be at $[\Phi_\mathrm{D1},\lambda_\mathrm{D1}]=[-35.8^{\circ},146.9^{\circ}]$ in Australia, and at $[\Phi_\mathrm{D2},\lambda_\mathrm{D2}]=[38.9^{\circ},-98.4^{\circ}]$ in North America (marked by $\times$ symbols). The location of the third telescope was chosen from the set of acceptable sites described in section \ref{section_map}.}\label{fig:AustraliaUSAC}
\end{figure}

As a first example for a $3$-telescope network (Example D), we chose one site at $[\Phi_\mathrm{D1},\lambda_\mathrm{D1}]=[-35.8^{\circ},146.9^{\circ}]$ in Australia, and one site at $[\Phi_\mathrm{D2},\lambda_\mathrm{D2}]=[38.9^{\circ},-98.4^{\circ}]$ in North America. The resulting plots of $\{I,D,R\}/\{I,D,R\}_\mathrm{max}$ ratios, given in percentages, are shown in figure \ref{fig:AustraliaUSA}. As shown in the $C/C_\mathrm{max}$ ratio plot in figure \ref{fig:AustraliaUSAC}, the optimal location for a third telescope in this network would either be in Central Africa or in the North-Central region of South America.

\begin{figure}
    \centering
\begin{tabular}{cc}
    \subfloat[]{\includegraphics[width=80mm]{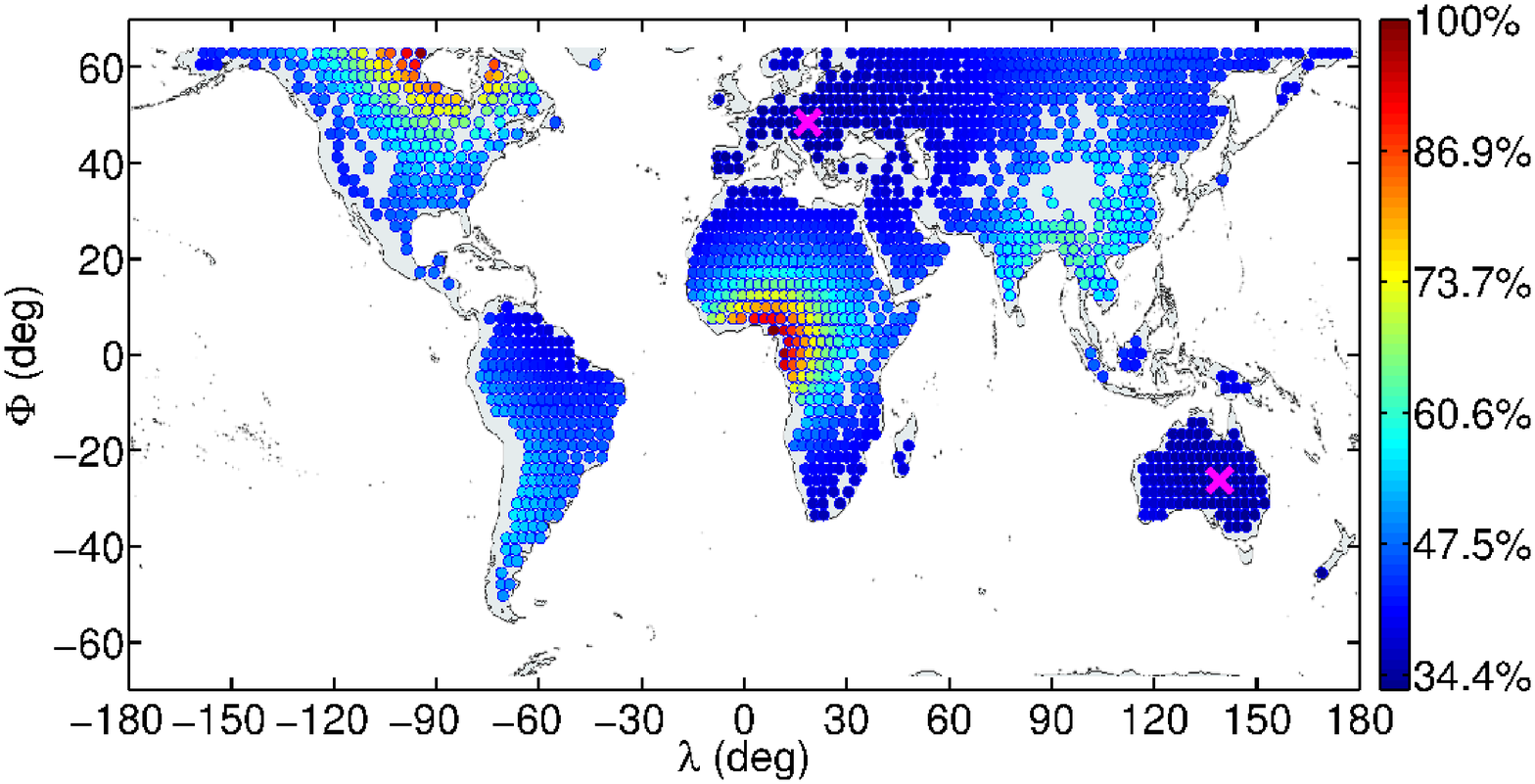}} & \hspace{-6mm} \subfloat[]{\includegraphics[width=80mm]{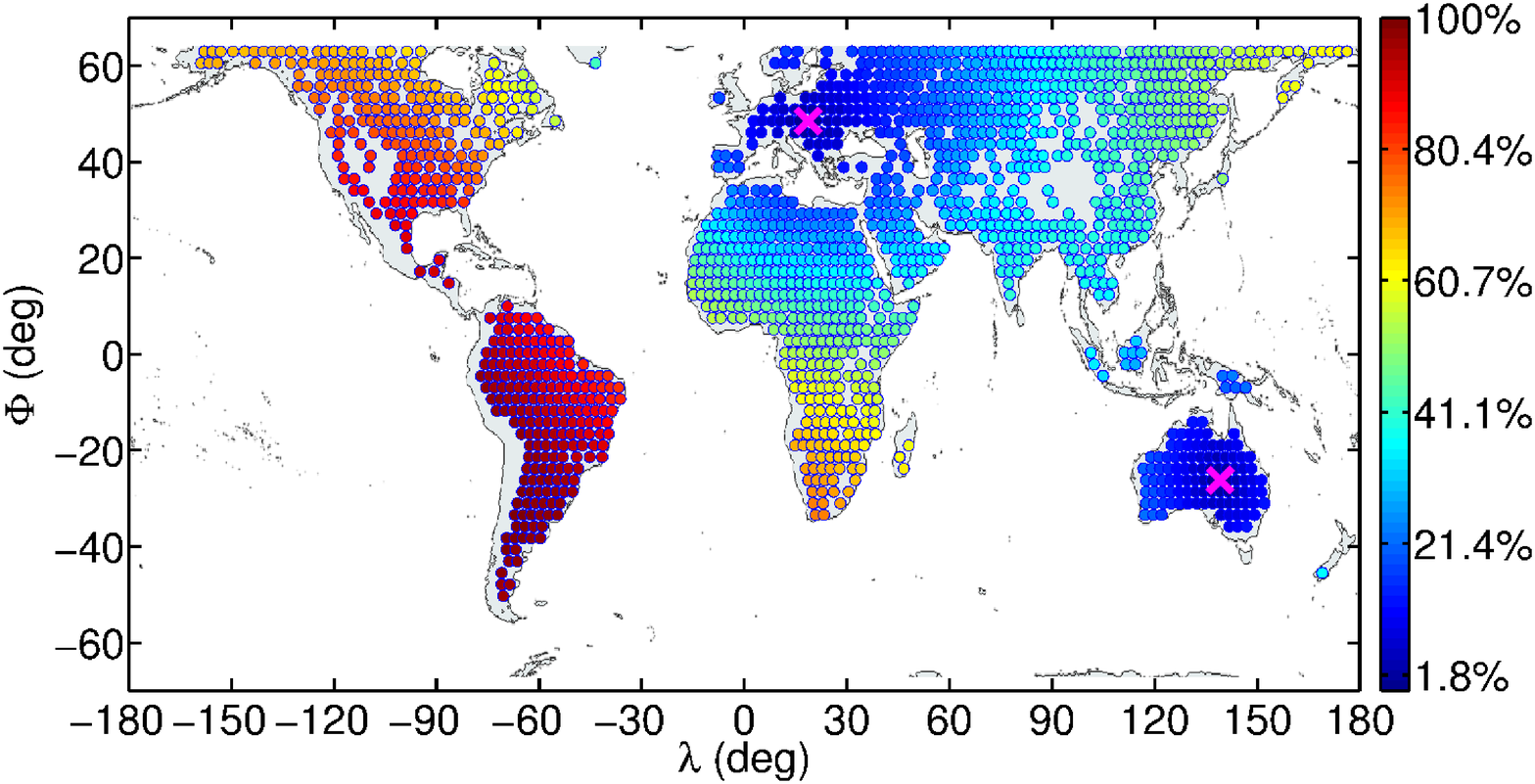}} \\
\end{tabular}
    \subfloat[]{\includegraphics[width=80mm]{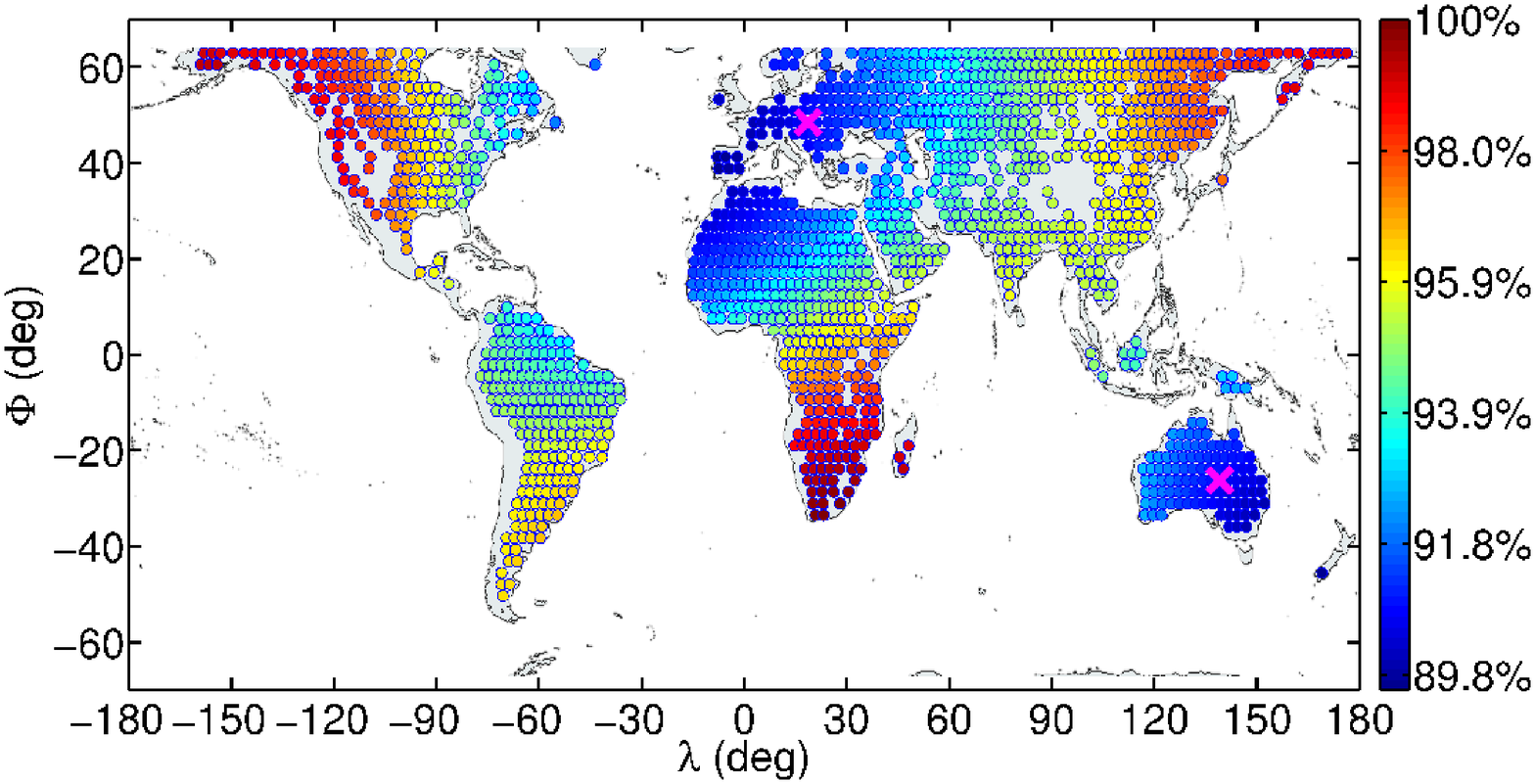}}
\caption{The colormaps of normalized metric values that we get for a network of three identical $\Delta$-telescopes in the case when one telescope is placed to $[\Phi_\mathrm{E1},\lambda_\mathrm{E1}]=[-26.2^{\circ},139.2^{\circ}]$ in central Australia, and one to $[\Phi_\mathrm{E2},\lambda_\mathrm{E2}]=[48.5^{\circ},18.7^{\circ}]$ in Europe (marked by $\times$ symbols). The location of the third telescope is chosen from the set of acceptable sites described in section \ref{section_map}. Each circle in the maps corresponds to one of the $1523$ site locations, while their colors represent the $I/I_\mathrm{max}$ (a), $D/D_\mathrm{max}$ (b), or $R/R_\mathrm{max}$ (c) values, given in percentages, where the ''max'' indices correspond to the highest values of the three metrics within the samples of $1523$ elements.
}\label{fig:AustraliaEurope}
\end{figure}

\begin{figure}
    \centering
    \includegraphics[width=120mm]{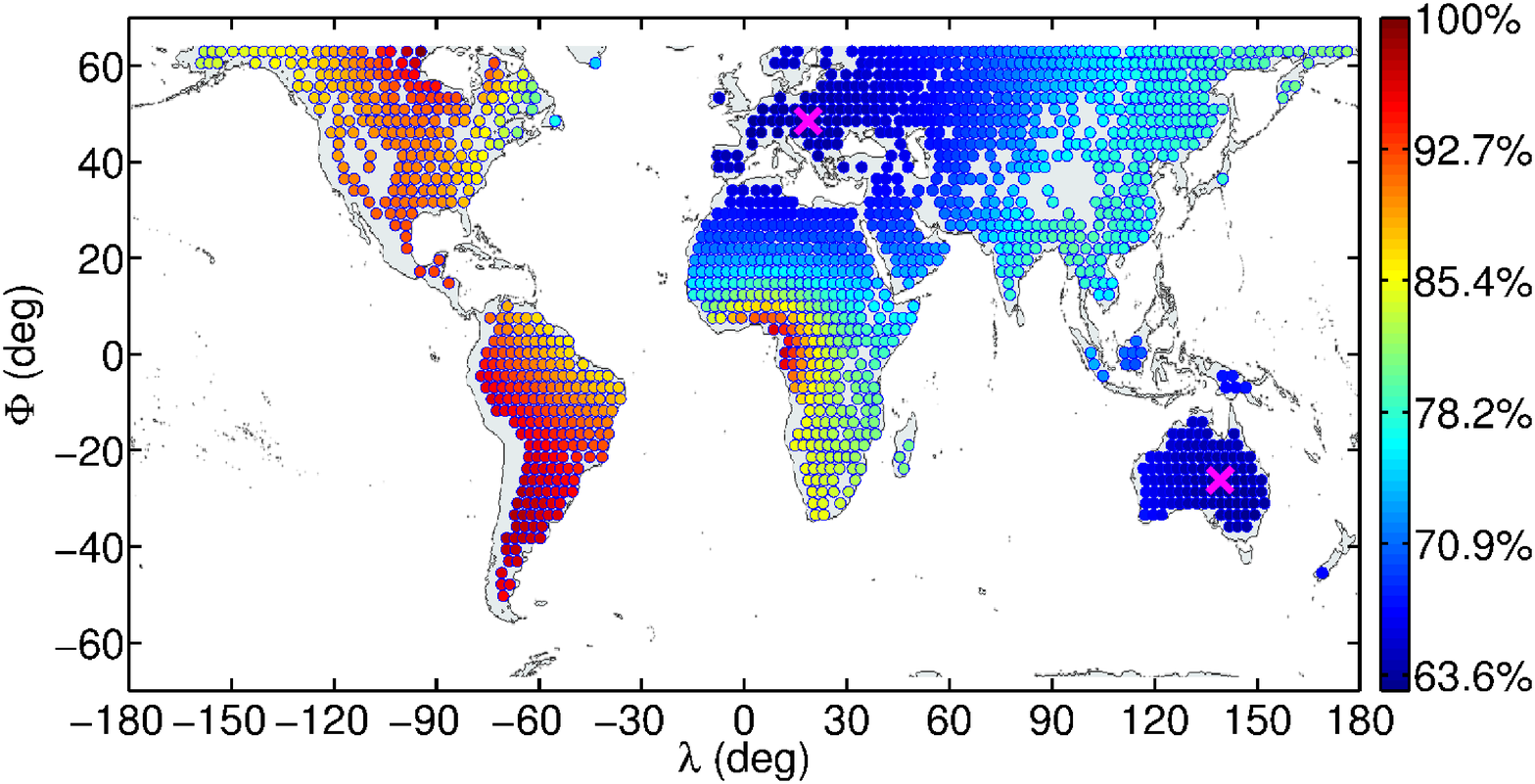}
\caption{The colormap of $C/C_\mathrm{max}$ values for $1523$ different network configurations of three identical $\Delta$-telescopes, where $C_\mathrm{max}$ is the highest value of the combined metric within the sample of $1523$ elements. The location of the first telescope in the network were chosen to be at $[\Phi_\mathrm{E1},\lambda_\mathrm{E1}]=[-26.2^{\circ},139.2^{\circ}]$ in central Australia, and at $[\Phi_\mathrm{E2},\lambda_\mathrm{E2}]=[48.5^{\circ},18.7^{\circ}]$ in Europe (marked by $\times$ symbols). The location of the third telescope was chosen from the set of acceptable sites described in section \ref{section_map}.}\label{fig:AustraliaEuropeC}
\end{figure}

As a second example for a $3$-telescope network (Example E), we chose one site at $[\Phi_\mathrm{E1},\lambda_\mathrm{E1}]=[-26.2^{\circ},139.2^{\circ}]$ in central Australia, and one at $[\Phi_\mathrm{E2},\lambda_\mathrm{E2}]=[48.5^{\circ},18.7^{\circ}]$ in Europe. The $\{I,D,R\}/\{I,D,R\}_\mathrm{max}$ ratios, given in percentages, are shown in figure \ref{fig:AustraliaEurope}, while the $C/C_\mathrm{max}$ ratios, also given in percentages, are shown in figure \ref{fig:AustraliaEuropeC}. The optimal locations for the third telescope in this case are in West Argentina and in West-Central Africa, based on the $C$ metric.

%% file: opt_India.tex
In this analysis, we used a similar approach as for the $\Delta$-telescope networks to find the optimal location and orientation for the proposed LIGO-India detector within a five-detector network with Advanced LIGO (Hanford), Advanced LIGO (Livingston), Advanced Virgo, and KAGRA. In this case, we had taken into account in the calculation of the $I$ and $R$ metric that the LIGO-India is proposed to be a single L-shaped interferometer with $4\ \mathrm{km}$ long arms, similarly to the Advanced LIGO interferometers. Also, we used the assumption again that the strain sensitivity of all the interferometers in the network are practically the same.

We first created a new set of acceptable sites for LIGO-India by focusing our filtered map (see section \ref{section_map}) to the Indian region and by resampling the map with a higher ($\sim 27\ \mathrm{km}$) resolution. The resulting discrete map of acceptable sites consisted of $5904$ different geographical locations, covering whole India and even extending beyond the country.

We manually chose $3$ site locations in India (in the Southern, in the North-Eastern and in the North-Western part of the country), and used each of them as a hypothetical site for LIGO-India. The orientation angle of the L-shaped interferometer was defined as the angle between the East direction and the bisector of the interferometer arms, measured counterclockwise. At each site, we calculated the $I$, $R$, $D$, and $C$ metrics for every orientation angle from $45^\circ$ to $135^\circ$ with a resolution of $0.1^\circ$. Note, that the $D$ metric is always independent from the orientation angle, and that the symmetry of projection of an incoming GW to the arms of an L-shaped interferometer makes it unnecessary to test any orientation angles below $45^\circ$ and above $135^\circ$. Based on the results, the optimal orientation angle for LIGO-India was robustly found to be between $57.7-58.9^\circ$ depending on the metric that we maximized ($+k\times 90^\circ$ due to symmetry reasons with $k$ being an integer).

In the case of site location optimization, instead of randomly choosing the site locations for LIGO-India, we carried out our analysis by testing all the $5904$ different sites. Using each acceptable site in our list as a hypothetical location for LIGO-India, we calculated the $I$, $D$, $R$, and $C$ metrics for the five-detector network, taking into account the known geographical positions and orientations of the Advanced LIGO Hanford and Livingston, Advanced Virgo, and KAGRA interferometers. The orientation angle for the LIGO-India interferometer was first set to be $45^\circ$ and colormaps of $I/I_\mathrm{max}$, $D/D_\mathrm{max}$, $R/R_\mathrm{max}$, and $C/C_\mathrm{max}$ values were produced, where the ''max'' indices corresponded to the highest value of the given metric within the sample of $5904$ elements.

\begin{figure}
    \centering
\begin{tabular}{cc}
    \subfloat[]{\includegraphics[width=70mm]{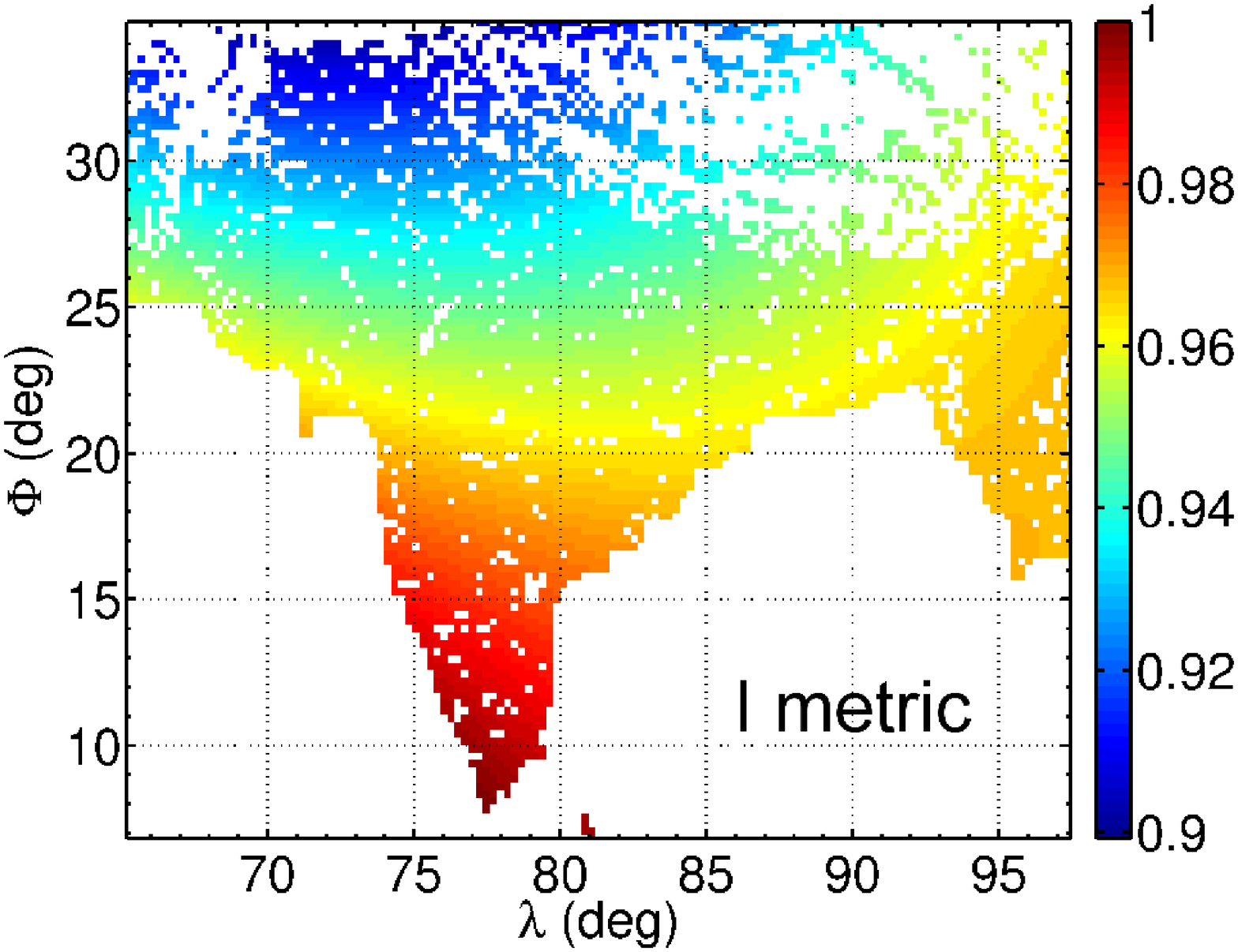}} & \hspace{-6mm} \subfloat[]{\includegraphics[width=70mm]{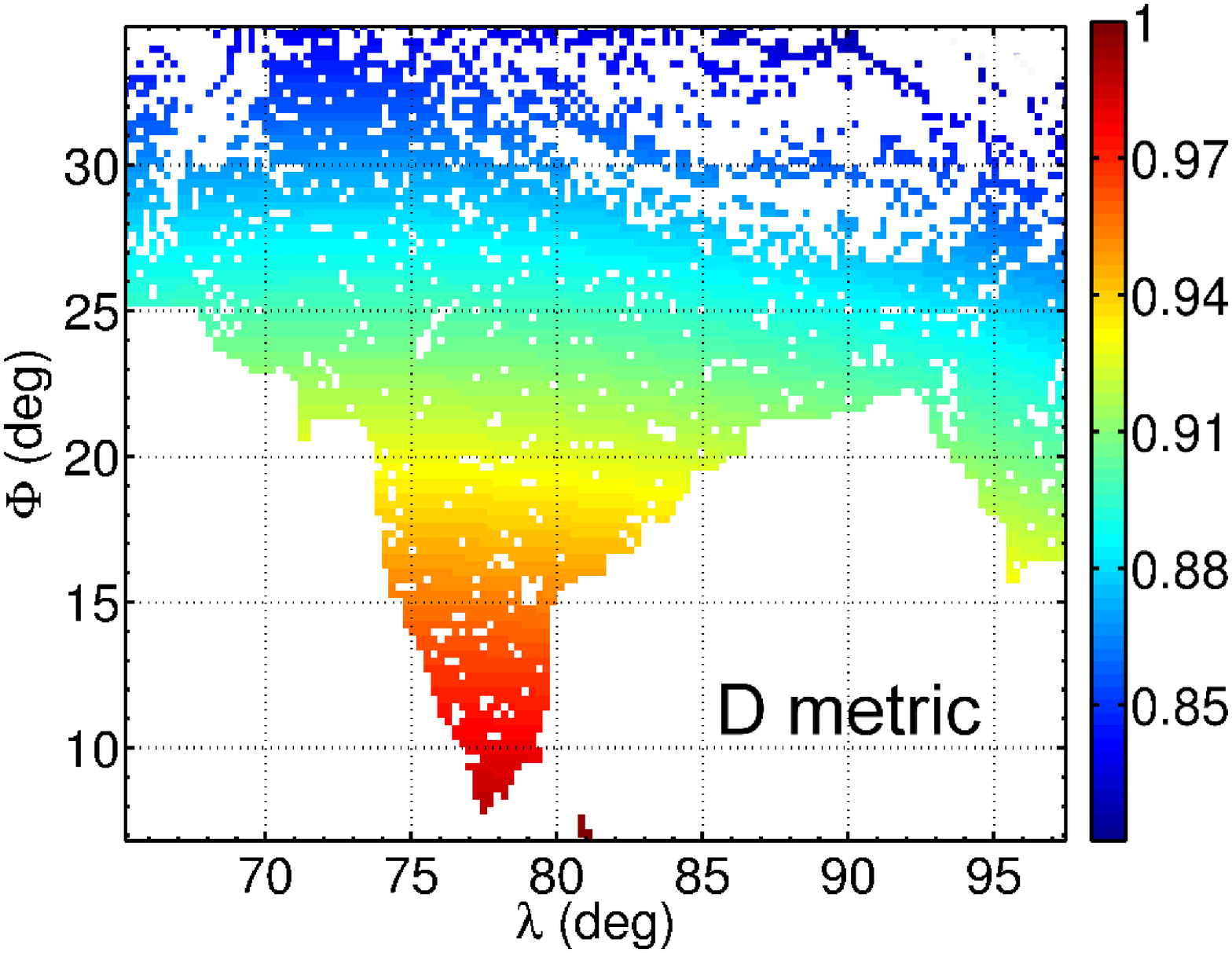}} \\
\end{tabular}
    \subfloat[]{\includegraphics[width=70mm]{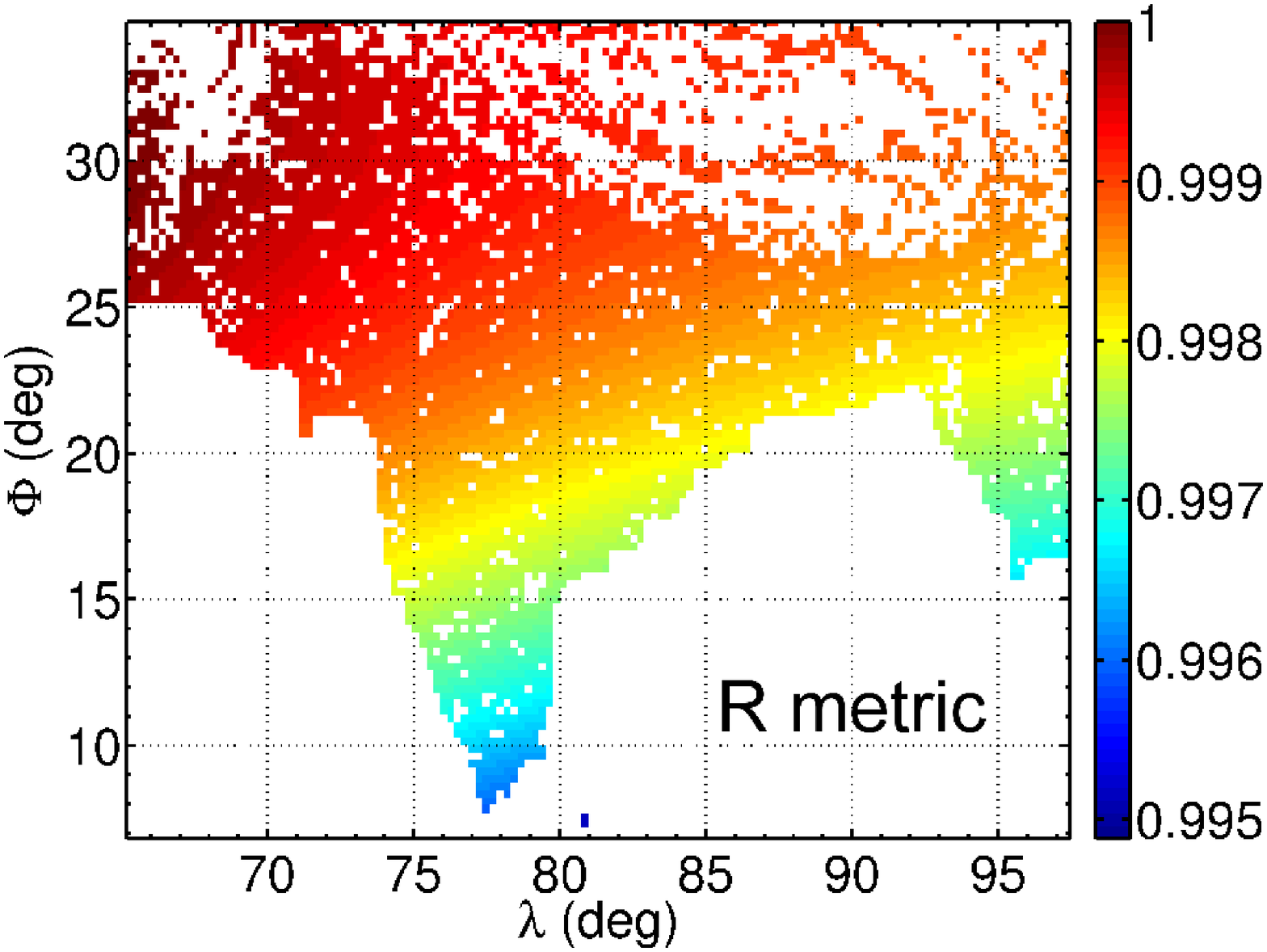}}
\caption{The colormaps of $I/I_\mathrm{max}$ (a), $D/D_\mathrm{max}$ (b), and $R/R_\mathrm{max}$ (c) values in case of placing the LIGO-India site at different acceptable site locations in the Indian subcontinent. The ''max'' indices correspond to the highest value of the different metrics within the samples of $5904$ elements. The orientation angle of the L-shaped interferometer (being the angle between the East direction and the bisector of the interferometer arms, measured counterclockwise) was optimized based on the $C$ metric, and was set to be $58.2^\circ$. The optimal orientation angle was found to be robust in terms of the site location within India.}\label{fig:Indias}
\end{figure}

\begin{figure}
    \centering
    \includegraphics[width=100mm]{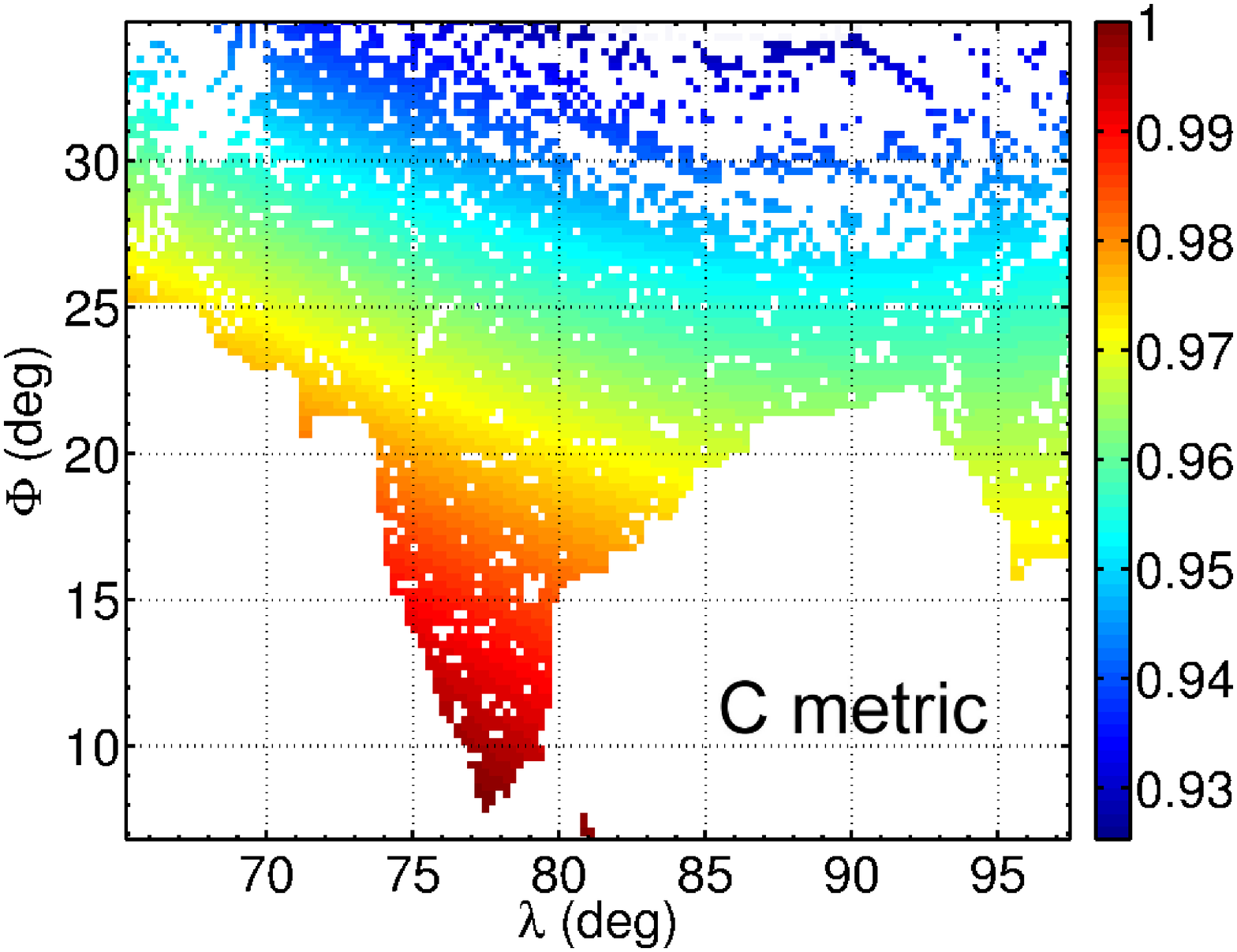}
\caption{This figure shows the same results as figure \ref{fig:Indias} but for the $C$ metric.}\label{fig:IndiaC}
\end{figure}

As we found that the $I$, $D$, and $C$ metric values are consistently higher in the Southern geographical regions, we repeated testing all the $5904$ sites using an orientation angle of $58.2^\circ$, which we had found to be the optimal angle for a LIGO-India placed to the Southern part of India, based on the combined metric, $C$. The results are visualized in figure \ref{fig:Indias} for the three individual metrics, and in figure \ref{fig:IndiaC} for the combined metric. Note, that testing the different site locations using a $45^\circ$ orientation angle resulted with very similar maps to the ones shown in figure \ref{fig:IndiaC}, which shows that the optimal site locations within India are also very robust to the orientation angle of the interferometer.

Based on the $I$, $D$, and the $C$ metrics, we found that the rule of thumb is that the five-detector network of LIGO-India, Advanced LIGO Hanford and Livingston, Advanced Virgo, and KAGRA, becomes more and more optimal by placing LIGO-India more to the South within the Indian subcontinent. Even though the $R$ metric favors the opposite geographical direction for the LIGO-India site, there is only a slight dependence of the network performance in terms of this metric.

\begin{figure}
    \centering
    \includegraphics[width=150mm]{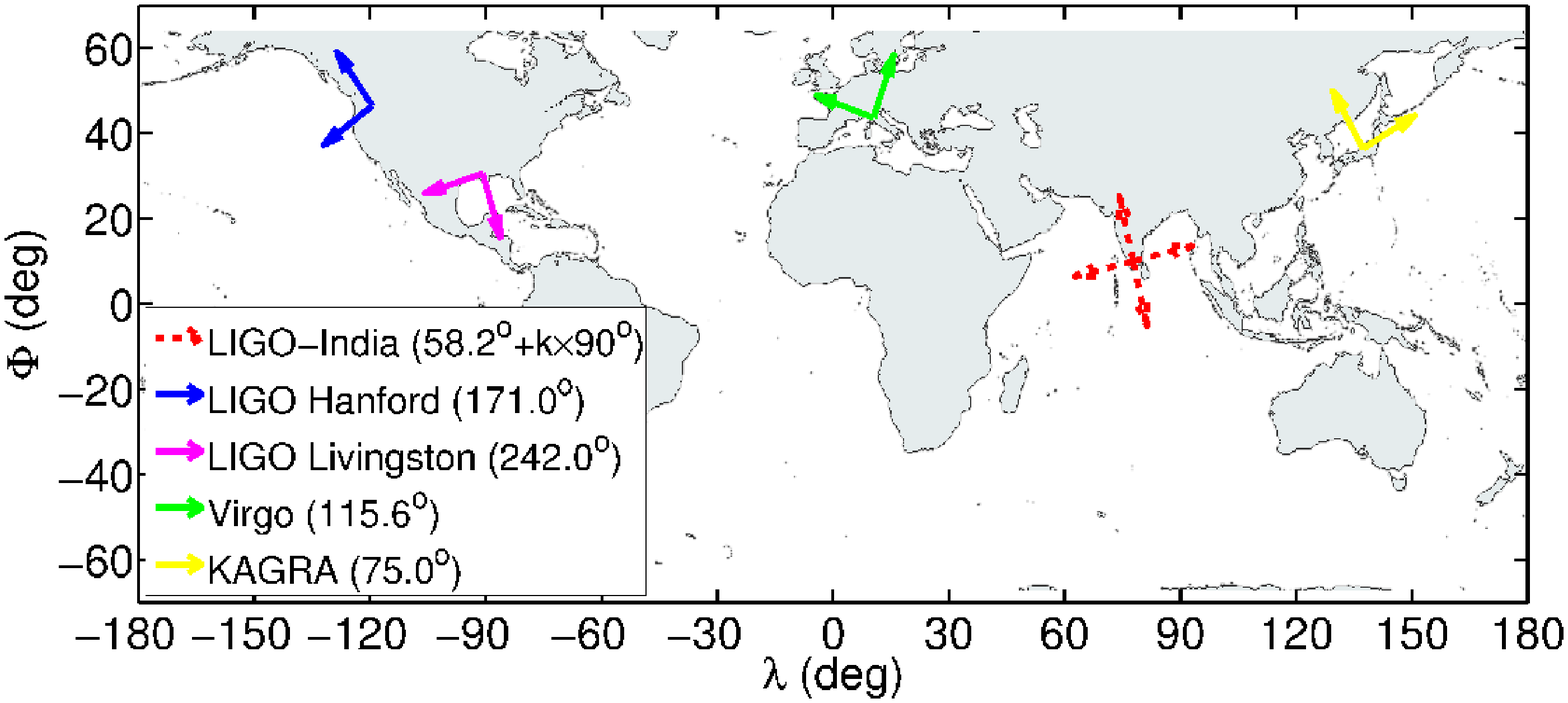}
\caption{A suggested example for an optimal network of second generation GW detectors including LIGO-India. In this example the LIGO-India is placed to $[\Phi=10.02^\circ,\lambda=77.76^\circ]$ with an orientation angle (being the angle between the East direction and the bisector of the interferometer arms, measured counterclockwise) of $58.2^\circ +k\times 90^\circ$, $k$ being an integer. The geographical positions and orientation angles of the other detectors in the network, $([\Phi,\lambda];\alpha)$, were set to $([46.4551^\circ,-119.41^\circ];171^\circ)$ for Advanced LIGO Hanford, $([30.56^\circ,-90.77^\circ];242^\circ)$ for Advanced LIGO Livingston, $([43.63^\circ,10.5^\circ];115.6^\circ)$ for Advanced Virgo, and $([36.42^\circ,137.3^\circ];75^\circ)$ for KAGRA.}\label{fig:IndiaOpt}
\end{figure}

\begin{figure}
    \centering
    \includegraphics[width=150mm]{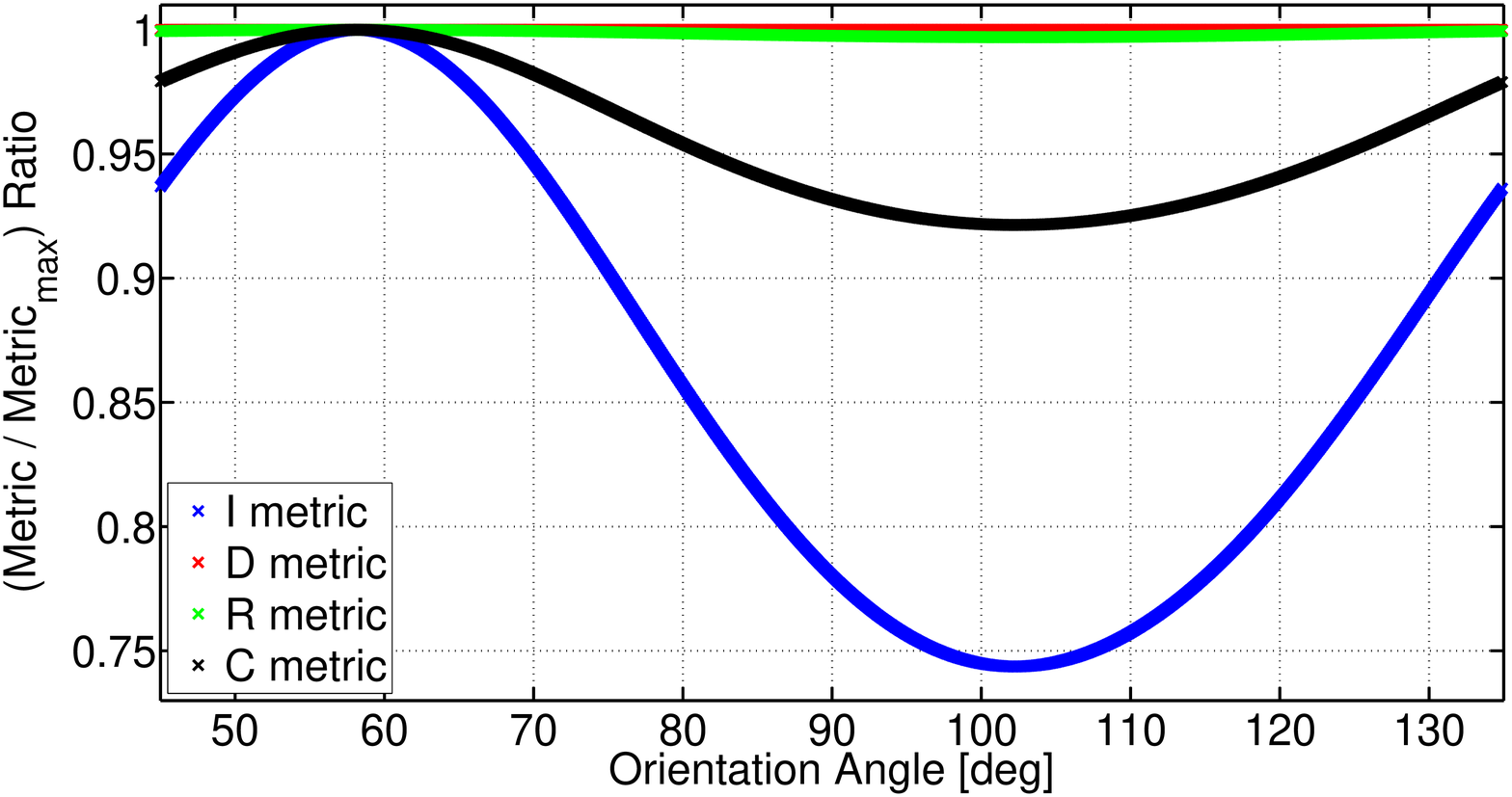}
\caption{The $I/I_\mathrm{max}$ (blue), $D/D_\mathrm{max}$ (red), $R/R_\mathrm{max}$ (green), and $C/C_\mathrm{max}$ (black) ratios (the ''max'' index corresponding to the highest values of the metrics within the sample) as functions of the detector orientation angle for a hypothetical LIGO-India detector at a geographical location $[\Phi=10.02^\circ,\lambda=77.76^\circ]$. Note, that the dominant metric in terms of optimizing the orientation angle is $I$, i.e. the capability of the network of reconstructing the polarization of a detected GW signal. This $I$ metric can suffer even a $\sim 25\%$ loss if the detector is not oriented optimally.}\label{fig:IndiaAngles}
\end{figure}

As a final result, we chose an example for an optimal configuration for LIGO-India to have the site location at $[\Phi=10.02^\circ,\lambda=77.76^\circ]$ and to have an orientation angle of $58.2^\circ +k\times 90^\circ$ (see figure \ref{fig:IndiaOpt}). The corresponding $I/I_\mathrm{max}$, $D/D_\mathrm{max}$, $R/R_\mathrm{max}$, and $C/C_\mathrm{max}$ ratios (the ''max'' index corresponding to the highest values of the metrics within the sample) as functions of the detector orientation angle are shown in figure \ref{fig:IndiaAngles}. Note, that the dominant metric in terms of optimizing the orientation angle of the LIGO-India detector, is $I$, i.e. the capability of the network of reconstructing the polarization of a detected GW signal. This $I$ metric can suffer even more than a $\sim 25\%$ loss if the detector is not oriented optimally.

%% file: Discussion.tex
In this section, we summarize the results presented in section \ref{section_positions} and \ref{section_India}, and suggest geographical regions where possible future $\Delta$-telescopes and the LIGO-India should be placed based on the $N$-telescope FoMs we defined in section \ref{section_maths}. We also suggest an optimal orientation for LIGO-India based on the same FoMs.

The FoMs we used to characterize and compare different configurations of $N$-telescope networks were the following (see section \ref{section_maths}): (a) capability of reconstructing the polarization of a detected GW signal (measured by the $I$ metric); (b) accuracy in localizing the GW source in the sky (measured by the $D$ metric); (c) accuracy in reconstructing the parameters of a standard binary source (measured by the $R$ metric). We also constructed a fourth metric (denoted by $C$) that combines the $I$, $D$, and $R$ metrics with practically equal weight (see section \ref{subsect:C}).

In section \ref{section_positions} we calculated the $I$, $D$, $R$, and $C$ maps in three example cases where the location of the first $\Delta$-telescope in a $2$-telescope network was set to be in Europe (see figure \ref{fig:2detFixedE} and \ref{fig:2detFixedCE}), North America (see figure \ref{fig:2detFixedU} and \ref{fig:2detFixedCU}), and India (see figure \ref{fig:2detFixedI} and \ref{fig:2detFixedCI}), respectively. The maps were always normalized with the highest value of the different metrics within the sample of $1524$ $I$, $D$, $R$, and $C$ values. We suggested a general method for choosing the optimal site for the first $\Delta$-telescope in the network, and for determining the corresponding set of optimal locations for a second instrument. We demonstrated this method with the combined metric, allowing a maximum of $-5\%$ loss in the $C/C_\mathrm{max}$ values of the different $2$-telescope networks. We found that for this tolerance criterion the optimal site locations for a second telescope form a $\pm 10^{\circ}$ annulus at an angular distance of $\sim 130^{\circ}$ from the location of the first instrument. Using this result we calculated the number of acceptable sites within such annuli associated to all the acceptable sites, and found that the highest number of optimal sites correspond to locations in Australia (see figure \ref{fig:OptNumMap}). The result of Australia being the optimal region for the first $\Delta$-telescope (assuming that one wants to preserve the most options in selecting an optimal site for a second telescope) proved to be very robust to a wide range of allowed maximum loss in $C/C_\mathrm{max}$.
 
We suggest that the site location for a second $\Delta$-telescope in the network should be chosen from the set of acceptable sites within the annulus of optimal sites associated to the location of the first site. We chose two example cases of $3$-telescope networks where the location of one telescope was chosen to be in Australia, and the second telescope was placed to North America (Example D; see figure \ref{fig:AustraliaUSA} and \ref{fig:AustraliaUSAC}), and to Europe (Example E; see figure \ref{fig:AustraliaEurope} and \ref{fig:AustraliaEuropeC}), respectively. The optimal sites for a third telescope in such network was found to be on Borneo island and in the North-Central region in South America for Example D, and in North Canada and in West-Central Africa for Example E.

Using an India map of acceptable telescope sites with $\sim 27\ \mathrm{km}$ resolution, we tested a total of $5904$ different site locations for LIGO-India, taking into account the Advanced LIGO Hanford and Livingston, Advanced Virgo, and KAGRA as further members of a five-detector network (see section \ref{section_India}). We found the optimal orientation angle of LIGO-India to be $\sim 58.2^\circ+k\times 90^\circ$ ($k$ being an integer) practically for all site locations within India, and based on the combined metric. Here, the orientation angle is defined as the angle between the East direction and the bisector of the arms in the L-shaped interferometer, measured counterclockwise.

Also, by using an optimally oriented configuration, we found the Southern-Central part of India to be the most optimal region for the site of LIGO-India (probably the South-Eastern part of the Karnataka region, or North-Western part of the Tamil Nadu region; see figure \ref{fig:IndiaC}). An example for an optimal location and orientation for LIGO-India is shown in Figure \ref{fig:IndiaOpt}. Comparing the results shown in figure \ref{fig:Indias} and figure \ref{fig:IndiaAngles}, we can also conclude that based on the network FoMs, setting the orientation angle of the LIGO-India detector has greater significance than choosing the site location of the detector within India. In the optimization of the site location, the dominant term will be the source localization accuracy of the network ($D$ metric), while in the optimization of the orientation angle, the dominant term will be the capability of the network of reconstructing the polarization of a detected GW signal ($I$ metric).

More efforts are required in the future to overcome the limitations of the work presented here. Taking a lot more exclusionary criteria into account, possibly with continuously variable acceptance instead of the present yes/no type grid, would allow constructing a much more detailed and accurate set of acceptable sites. Applying public crowdsourcing in setting up a list of exclusionary criteria and/or inaccessible geographic areas would result with such a detailed map of possible constructions sites that probably no group of researchers would be able to achieve.

Additional FoMs can easily be added to the optimization process depending on the aim of various searches from GW transients to the stochastic GW background. One such FoM could be the expected duty cycle of the network or a sub-network of a given number of telescopes, as proposed by \cite{BFS11}. This FoM and potentially others could allow not only comparing different configurations of the same $N$-telescope network (as presented here) but also networks with different number of telescopes, leading to optimization of the number of telescopes in the network as well.

The $I$, $D$, and $R$ metrics have already been used in similar forms (without all-sky averaging) in previous works (see \cite{SK11}, \cite{SF11}, and \cite{PA09}). These metrics can unquestionably be used to find the {\it optimal} configurations of $N$-telescope networks for all given $N$s, however any of them might need more elaboration if one wants to use them to make quantitative comparisons between optimal and sub-optimal configurations, or between networks of different numbers of telescopes. Also, one can use a different weighing of individual metrics when constructing the combined ($C$) metric depending on the specific preferences one would like to follow, which might also be a subject to change with time as we move closer to the realization of future GW telescopes. These are all beyond the scope of this paper and should be investigated in a future follow-up paper.

Finally, the optimization of $N$-telescope networks could be based on regular or Markov chain Monte Carlo simulations using the set of acceptable sites and metrics. This would allow a more general testing of networks of $N\geq 3$ $\Delta$-telescopes. Also, as we pointed out in section \ref{section_maths}, one could take into account the distribution of mass in the local Universe (e.g. using galaxy catalogs), and optimize the telescope network by maximizing the observational time with the highest directional sensitivity towards certain sky directions. We also leave these to be the scope of a future paper.

%% file: Acknowledgments.tex
This paper was reviewed by the LIGO Scientific Collaboration under LIGO Document $\mathrm{P1200175}$. We'd like to thank Imre Bartos, Brian Dawes, Riccardo DeSalvo, Marco Drago, Sergei Klimenko, Gabriele Vedovato, Linqing Wen, and Stan Whitcomb for their valuable comments on the manuscript. The authors are grateful for the support of the LIGO-Virgo Collaboration, and Columbia University in the City of New York. This work has been supported by the United States National Science Foundation under cooperative agreement PHY-$0847182$, the Science and Technology Facilities Council of the United Kingdom, and the Scottish Universities Physics Alliance.